\documentclass[5p,twocolumn,10pt,times]{elsarticle}
\usepackage{amsmath}
\usepackage{hyperref}
\usepackage{esvect}
\usepackage{multirow}

\usepackage{longtable}
\usepackage{booktabs}
\usepackage{ctable}
\usepackage{soul}
\usepackage{ulem}
\usepackage{wrapfig}

\newcommand{\rev}[2]{#2}

\usepackage{color}

\begin{document}
\baselineskip11pt

\begin{frontmatter}

\title{Exceptional Mechanical Performance by Spatial Printing with Continuous Fiber:~\rev{}{Curved Slicing, Toolpath Generation and Physical Verification}}

\author{Guoxin Fang\textsuperscript{1}}
\author{Tianyu Zhang\textsuperscript{1}}
\author{Yuming Huang\textsuperscript{1}}
\author{Zhizhou Zhang\textsuperscript{1}}
\author{Kunal Masania\textsuperscript{2}}
\author{Charlie C.L. Wang\textsuperscript{1,*}}

\address{\textsuperscript{1}School of Engineering, The University of Manchester, United Kingdom \\
\textsuperscript{2}Faculty of Aerospace Engineering, Delft University of Technology, the Netherlands}

\cortext[cor1]{Corresponding author. E-mail: changling.wang@manchester.ac.uk}

\begin{abstract} 
This work explores a spatial printing method to fabricate continuous fiber-reinforced thermoplastic composites (CFRTPCs), which can achieve exceptional mechanical performance. For models giving complex 3D stress distribution under loads, typical planar-layer based fiber placement usually fails to provide sufficient reinforcement due to their orientations being constrained to planes. The effectiveness of fiber reinforcement could be maximized by using multi-axis additive manufacturing (MAAM) to better control the orientation of continuous fibers in 3D-printed composites. 
Here, we propose a computational approach to generate 3D toolpaths that satisfy two major reinforcement objectives: 1) following the maximal stress directions in critical regions and 2) connecting multiple load-bearing regions by continuous fibers. 
Principal stress lines are first extracted in an input solid model to identify critical regions. Curved layers aligned with maximal stresses in these critical regions are generated by computing an optimized scalar field and extracting its iso-surfaces. Then, topological analysis and operations are applied to each curved layer to generate a computational domain that preserves fiber continuity between load-bearing regions. Lastly, continuous fiber toolpaths aligned with maximal stresses are generated on each surface layer by computing an optimized scalar field and extracting its iso-curves. A hardware system with dual robotic arms is employed to conduct the physical MAAM tasks depositing polymer or fiber reinforced polymer composite materials by applying a force normal to the extrusion plane to aid consolidation. When comparing to planar-layer based printing results in tension, up to $644\%$ \rev{breaking forces}{failure load} and $240\%$ stiffness are observed on shapes fabricated by our spatial printing method. We demonstrate the versatility of our approach through various complex load cases which demonstrate their successful implementation of continuous fiber printing in 3D.
\end{abstract}


\end{frontmatter}

\section{Introduction}\label{sec:intro}

\begin{figure*}[t]
 \centering 
\includegraphics[width=\linewidth]{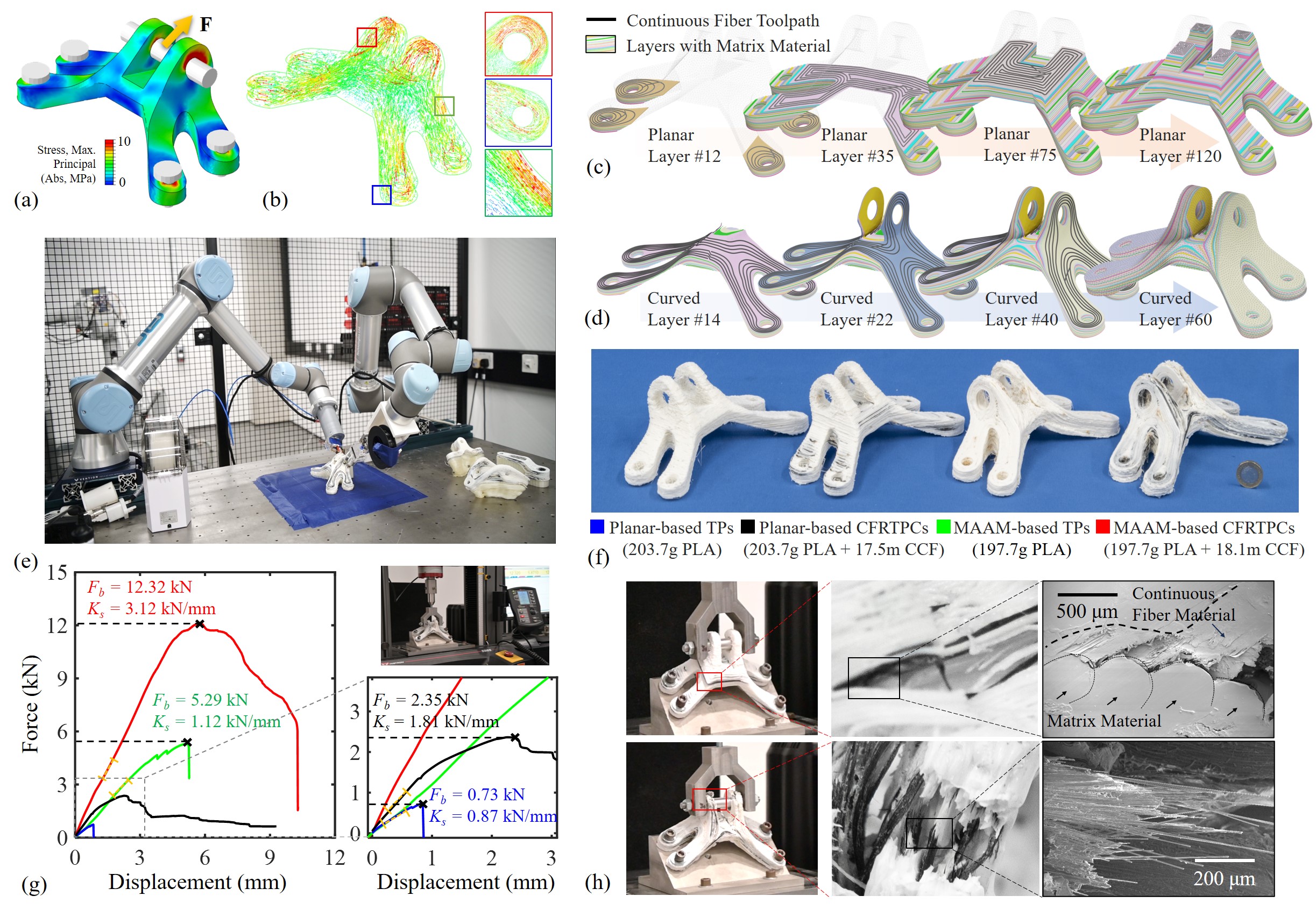}
\caption{
(a) Loading condition and stress distribution for a GE-Bracket model generated by topology optimization.
(b) Principal stress flow highlighting the critical stress directions in some localized regions.
(c) Toolpaths generated by planar-based layer slicing are misaligned with the stress distribution in 3D space and also fail to connect load-bearing regions -- both limit the effectiveness of fiber reinforcement. 
(d) By utilizing the computational framework proposed in this study, curved layers and 3D fiber toolpaths are generated to effectively reinforce 3D printed model -- note that the load-bearing regions are effectively connected by fiber toolpaths on the same layer.
(e) Hardware setup with dual robotic arms are employed to fabricate CFRTPCs with fibers aligned in 3D space accordingly.
(f) Models fabricated using different strategies.
(g) Tensile tests reveal significant improvements in both the \rev{breaking force}{failure load} and the stiffness for the CFRTPCs produced by our fiber printing with CCF. 
(h) Material characterization with zoom-in views (middle column) and scanning electron microscope (SEM) images of the breaking regions (right column). \rev{ which shows the material failure has changed from the delamination between fiber and matrix material (observed on the planar printing result as shown in the top row) into the fiber fracture (observed on the spatial printing result as shown in the bottom row).}{}
}
\label{fig:teaser}
\end{figure*}

Recent advancements in additive manufacturing (AM) have opened new opportunities to fabricate components with complex geometries and achieve high stiffness with excellent strength-to-weight ratio~\cite{kabir2020critical, gantenbein2018three, gantenbein2021spin}. Of these approaches, 3D printing of continuous fiber-reinforced thermoplastic composites (CFRTPCs), \rev{}{utilizing either in-nozzle impregnation~\cite{Matsuzaki2016_SciReport} or out-of-nozzle impregnation~\cite{DICKSON2017146} strategies,} has emerged as an alternative to conventional time-consuming and labor-intensive industrial molding processes~\cite{mallick2007fiber}. \rev{}{To optimize the performance of 3D printed CFRPTCs, studies have been made by optimizing printing parameters (e.g., the temperature and the layer height~\cite{Tian16_composite, heidari2019mechanical}) and developing post-processing techniques~\cite{Omuro2017, Nekoda21_AdditManu}.} \rev{Since }{On the other hand,} continuous fibers demonstrate maximum mechanical properties along the fibers' axial direction, the anisotropic mechanical strength \rev{}{that is controlled by printing toolpaths} can be \rev{well controlled}{optimized} to achieve a better reinforcement. Load-dependent fiber toolpaths have been investigated in studies~\cite{prashanth2017fiber,  Narasimha20_SciReport, liu2023stress}. \rev{that can significantly enhance the mechanical performance of 3D printed models. Additionally}{Meanwhile,} objectives such as optimizing fiber volume fractions~\cite{saeed2022characterization}, ensuring toolpath continuity~\cite{zhang2023graph}, and managing the turning angle of fibers~\cite{huang2023turning, zhang2021fibre} have been key considerations in generating toolpaths for 3D printing CFRTPCs. \rev{}{However, these studies primarily focused on models with in-plane stress distribution. The generation of fiber toolpaths for models with 3D stress flow (as shown in Fig.~\ref{fig:teaser}) was less explored.}

\rev{Previous studies of 3D printed CFRTPCs primarily employ}{Planar-layered based AM processes are commonly used to fabricate CFRPTCs in both the academic research~\cite{Li2020_CompositeB, chen2022field} and the industrial solutions (e.g., MarkForged~\cite{Eiger}, Anisoprint~\cite{Anisoprint}, and 9T Lab~\cite{9TLab}).}~These processes integrate continuous fibers with matrix material in a 'two-and-a-half-dimensional' (2.5D) manner~\cite{gao2015status}, which significantly reduces the complexity of the toolpath and the cost of the machine. \rev{has therefore been widely adopted in the commercial mass production of CFRTPCs~\cite{Eiger}.}{} Prior research has observed that the mechanical strength of 3D printed CFRTPCs can be notably enhanced when 
\begin{enumerate}
    \item Continuous fibers are placed along the directions of maximal principal stresses~\cite{chen2022field, heitkamp2023stress};
    
    \item Loops of continuous fibers are formed to connect the multiple load-bearing regions~\cite{Li2020_CompositeB, Sugiyama20_CST}.
\end{enumerate}
\rev{}{For models with loading conditions that result in 3D stress distribution,} meeting these requirements becomes challenging by planar-layer based AM, as this method limits fiber placement in the third dimension. 
This limitation is evident in the 2.5D slicing and toolpath generation for the GE-bracket model (Fig.~\ref{fig:teaser}(c)), where critical load-bearing areas, such as the holes with bolted joints, are not effectively interconnected.
Additionally, the in-plane placement of fibers fails to align with the \rev{}{principal} stress directions. This misalignment\rev{, along with the separation of critical regions,}{} leads to substantial peeling forces and potentially causes structural fractures like delamination between fibers and matrix material, as shown in the top row of Fig.~\ref{fig:teaser}(h).
\rev{}{In such scenarios, the effectiveness of fiber reinforcement can be significantly reduced~\cite{heitkamp2023stress}. To tackle these challenges caused by complex 3D stress distributions, we investigate a spatial printing method for CFRTPCs and develop a new toolpath generation algorithm that guides continuous fiber alignment in three-dimensional space.}

\vspace{5px}

\rev{}{Hardware systems of multi-axis additive manufacturing (MAAM)~\cite{urhal2019robot, zhang2021singularity}} \rev{providing a}{provide} higher degrees-of-freedom (DOF) for material accumulation and \rev{enabling}{enable} spatial printing. The MAAM process \rev{facilities}{can help to realize} advanced \rev{}{manufacturing} objectives, including support-structure elimination~\cite{dai2018support} and \rev{improve surface finishing quality}{surface-quality improvement}~\cite{huang2021conformal,etienne2019curvislicer}. \rev{}{Research efforts have started to employ MAAM in manufacturing with functional materials,} \rev{This includes hi-performance}{including high-performance} polymers~\cite{guidetti2023stress}, metals~\cite{bhatt2022robot}, and conductive filaments~\cite{hong20235}. In our previous work~\cite{fang2020reinforced}, the application of MAAM for reinforcing thermoplastic materials, such as polylactic acid, has been explored. \rev{}{However, the development of 3D toolpaths for the continuous fiber is still in its early stages of investigation. The challenge lies in the computational complexity involved in exploring the high dimensional design space, where} multiple objectives need to be \rev{taken into account, including aligning fibers with stress directions and ensuring toolpath continuity - as highlighted in the previous paragraph}{considered -- such as stress directions, toolpath continuity}. The existing method of 2.5D slicing and toolpath generation for continuous fiber, while effective in certain scenarios, cannot be directly extended to 3D cases. Moreover, the approach by Zhang~\textit{et al}~\cite{zhang2023robot} which can generate conformal fiber toolpaths on single-layer surface models, faces limitations when applied to solid models.

\rev{}{In this paper, we propose a field-based computational pipeline that can generate spatial fiber toolpaths optimized according to the requirements of fiber reinforcement.} Different from layer-free methods that directly generate printing toolpaths by decomposing the 3D space~\cite{retsin2016discrete, shembekar2019generating,cam2023fluid}, our method first generates curved layers (for both the fibers and the matrix material) and then computes topology preserved toolpaths (for continuous fibers). The toolpath computed with this strategy naturally generates the sequence of manufacturing to be implemented by MAAM~\cite{xu2019curved,bi2023strength}. Our computation is based on the stress analysis of the model with Finite Element Analysis (FEA), where a principal stress line (PSL) guided algorithm is invited to identify critical regions to guide curved layer slicing. On each curved layer, unlike \cite{fang2020reinforced}, we apply topological analysis and operations to generate a computational domain that preserves the continuous fiber connection between load-bearing regions. The computation result for the GE-Bracket model is illustrated in Fig.~\ref{fig:teaser}(d), which effectively meets the requirements for optimal fiber reinforcement - the generated toolpaths maintain continuity across each curved working surface and align with the 3D stress distribution in critical regions.

\rev{To verify the effectiveness of the proposed MAAM with CFRPTCs using designed spatial toolpaths.}{We have conducted physical experiments to verify the effectiveness of our spatial toolpaths for printing CFRPTCs.} The hardware \rev{setup is depicted}{system is shown} in Fig. 1(e), which contains \rev{a dual robotic arm MAAM system}{dual robotic arms}.  
\rev{The GE-Bracket model has been fabricated using different methods (see Fig.~\ref{fig:teaser}(f)).}{We fabricated models with 3D stress distribution using this MAAM system and compared their performance with models created using planar toolpaths and conventional AM processes (see Fig.~\ref{fig:teaser}(f) for an example of the GE-Bracket model).} Similar lengths of continuous carbon fibers (CCF) are employed in planar-based CFRTPCs (using \rev{zigzag}{contour-parallel} toolpath for CCF, as shown in Fig.~\ref{fig:teaser}(c)) and MAAM-based CFRTPCs (using our optimized toolpath for CCF). Planar-based layers and curved layers are also used to fabricate models by only using the matrix material (i.e., PLA). These are denoted as Planar-based TPs and MAAM-based TPs, respectively. \rev{}{A similar amount of PLA filaments is employed to fabricate these models. All layers of matrix materials are fabricated using toolpaths with a contour-parallel pattern 
} To ensure a fair comparison, curved layers \rev{}{of MAAM-based TPs} are also generated by our method. The planar-layers are generated by CURA \cite{Cura}.  

\rev{Material testing revealed that the GE-bracket model, fabricated with our spatial toolpaths, exhibited significant enhancements in failure load and stiffness – increases of $417.6\%$ and $72.4\%$, respectively, compared to planar-based CFRTPCs.}
{Mechanical tests revealed that the GE-bracket model fabricated with our spatial toolpaths has its failure load and stiffness increased by $417.6\%$ and $72.4\%$ respectively, comparing to the planar-based CFRTPC.}
It can also be observed from the bottom row of Fig.~\ref{fig:teaser}(h) that the breakage of \rev{the 3D printed CFRTPC}{the CFRTPC printed} by our spatial toolpath is due to fiber fracture. 
\rev{The structure failure occurs}{We can also observe structural failure} in the load-bearing region (e.g., bolt joints) where the highest stresses are presented. This also indicates that the reinforcement using our spatial toolpaths has aligned the strongest orientations of fibers along the directions of maximal stresses\rev{}{.}

To the best of our knowledge, this is the first work that explores the problem of MAAM for CFRTPCs and can compute optimized 3D continuous fiber toolpaths for models with complex geometry. 
The technical contributions of our computational pipeline are summarized as follows:
\begin{itemize}
\item An improved formulation for computing an optimized scalar field that generates curved layers aligned with maximal stresses. This method utilizes Principal Stress Lines (PSL) to continuously identify critical regions and effectively weight elements.


\item A novel method based on topology analysis for generating fiber toolpaths on each curved layer, which continuously connects the boundaries of load-bearing regions while complying with maximal stresses in critical regions.
%
\end{itemize}
Material characterization and tensile tests have been conducted to analyze the resultant CFRTPCs. It practically demonstrates the effectiveness of spatial fiber printing, which can achieve exceptional mechanical performance. 

The rest of the paper is organized as follows. \rev{Sec.~\ref{Sec:RelatedWork} reviews the recent research in 3D printing for CFRTPCs and spatial toolpath generation. }{We first present the} computational pipeline for spatial fiber toolpath generation in Sec.~\ref{Sec:ComputationalPipeline}, followed by discussing the computational results, the details of fabrication experiments, and the results of mechanical tests in Sec.~\ref{Sec:Result}. Finally, the paper concludes in Sec.~\ref{Sec:Conclusion}. 
\section{Non-planar Slicer and Toolpath Generator for CFRTPCs}~\label{Sec:ComputationalPipeline}
\begin{figure*}[t]
\centering 
\includegraphics[width=0.8\linewidth]{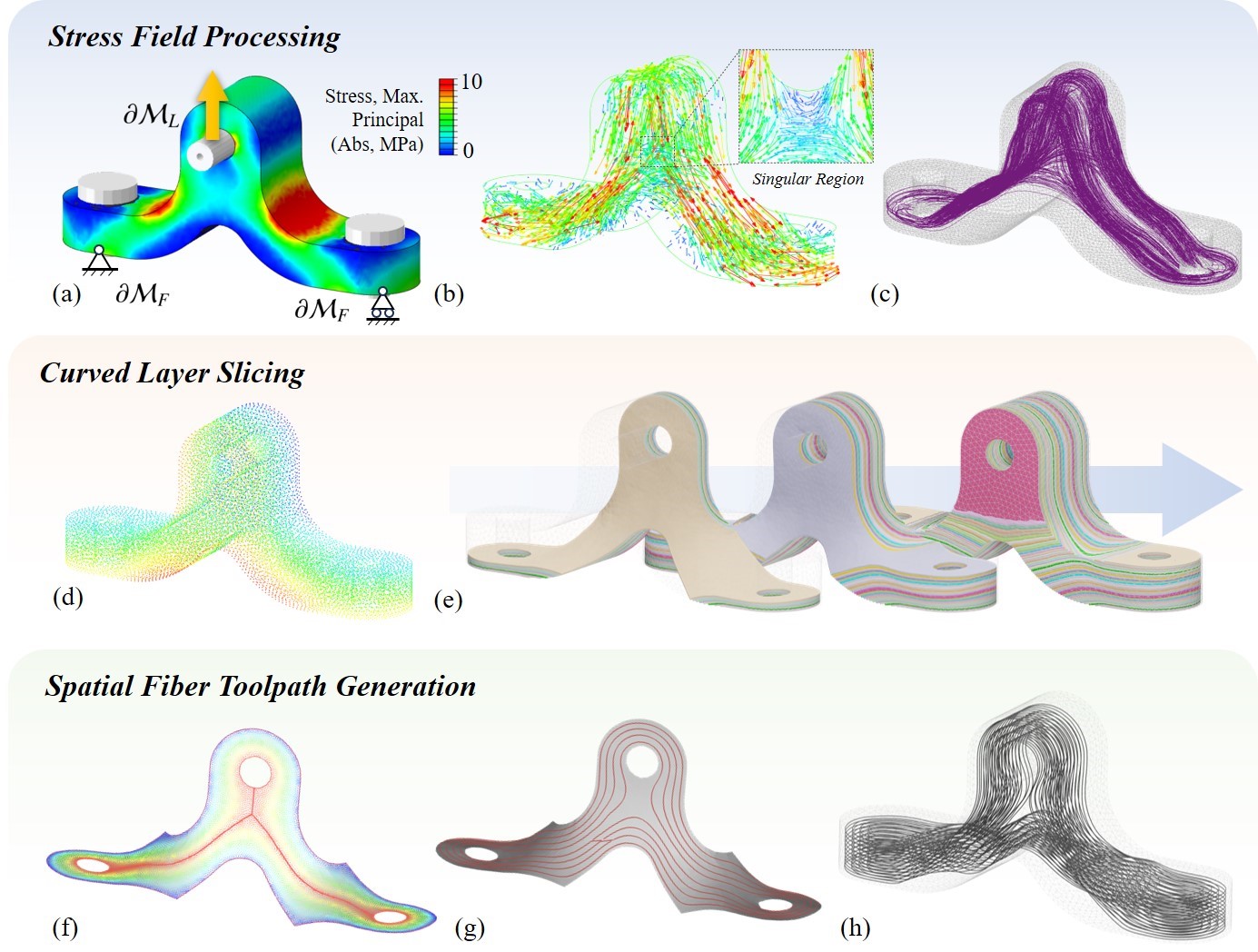}
\caption{The computational pipeline for generating spatial continuous fiber toolpaths, taking into account both stress orientation and geometric continuity to reinforce critical regions. 
(a) Given the target model $\mathcal{M}$ with applied boundary conditions as $\{ \partial\mathcal{M}_{F}, \partial\mathcal{M}_{L} \}$, principal stress distribution (represented as a vector field $\mathcal{V}$ in (b)) are computed by FEA. 
(c) A set of principal stress lines $\{ \mathcal{L} \}$ is traced to guide the computation of (d) the scalar field $G$ by optimization. This field subsequently guides the creation of (e) a sequence of curved layers $\{ \mathcal{S} \}$ as iso-surfaces of $G$. A similar field-based method is applied to each curved layer to generate continuous fiber toolpaths in 3D space. 
(f) Geometry analysis is first applied to segment $\mathcal{S}$ at the singularity region, and a scalar field $P$ is computed by optimizing stress-aware and toolpath-continuity objectives. 
(g) Continuous fiber toolpaths $\mathcal{T}$ are generated to effectively connect load-bearing regions for realizing optimized reinforcement. 
(h) The computing result is a set $\{\mathcal{T}\}$ of fiber toolpaths with sequence information -- note that only a subset of the toolpaths are displayed here for the purpose of a better visualization.}
\label{fig:framework}
\end{figure*}

In this section, we present the computational framework for generating continuous spatial toolpaths that optimize the mechanical properties of CFRTPCs. The computed toolpaths ensure stress alignment in key areas and maintain continuity to connect load-bearing regions. The computational pipeline includes the steps of stress field processing, field-guided curved-layer slicing, and spatial toolpath generation, as illustrated in Fig.~\ref{fig:framework}.

\subsection{Algorithm Overview}
The input of our computational pipeline is the target model shape represented by a volumetric mesh $\mathcal{M}$ which contains a set of tetrahedral elements $\{e\}$. Given the boundary condition (including the region with fixture $\partial\mathcal{M}_F$ and the region with external loads $\partial\mathcal{M}_L$), the stress distribution in $\mathcal{M}$ can be computed by finite element analysis (FEA). Specifically, the maximum principal stress directions, where each \rev{}{$\Vec{\sigma}_{max}(e)$} represents the most critical stress direction inside an element $e$, form a bidirectional stress field $\mathcal{V}=\{\pm \Vec{\sigma}_{max}(e)\}$ as illustrated in Fig.~\ref{fig:framework}(b). Since this stress field can have irregularities and inconsistencies in singular regions, we seek the help of the traced continuous principal stress lines (PSLs) $\{\mathcal{L}\}$ to process $\mathcal{V}$ and create a more harmonic stress distribution.
Details are presented in Sec.~\ref{subsec:stressFieldProcessing}.  

After that, the processed stress field is utilized to guide the computation of a scalar field $G$ by solving an optimization problem that minimizes design and manufacturing objectives related to fiber reinforcement. \rev{}{The gradient of this scalar field $\nabla G(e)$ governs the directions of material growth. The magnitude of this gradient, $\|\nabla G(e)\|$, determines the growth speed thus influences the variation of layer thickness.} Specifically, two objectives are considered: 1) the stress following term as $E_{sf}$ and 2) the continuity preservation term as $E_{cp}$. Moreover, we incorporate an objective for \rev{harmonic field $E_{hf}$}{controlling the compatibility of gradients $\nabla G(e)$ -- denoted by $E_{cg}$} \rev{- is also added serving}{to serve} as a regularization in the optimization pipeline. Details are presented in Sec.~\ref{subsec:fieldComputing}. Since a scalar value $g(v)$ is assigned to each vertex $v \in \mathcal{M}$, the iso-surfaces $\{ \mathcal{S}\}$ of $G$ can be extracted on $\mathcal{M}$ as curved triangle meshes. These mesh surfaces will later serve as the curved printing layers, allowing continuous toolpaths to be placed in the most critical direction to prevent delamination between layers. Additionally, load-bearing regions (e.g., bolt joints in Fig.\ref{fig:framework}) are geometrically connected on these curved surfaces. 

Finally, continuous fiber toolpaths are computed on these curved layers with the method explained in Sec.~\ref{subsec:toolpathGeneration}. A field-based approach is again applied, which first computes an optimized scalar field $P$ that considers three objectives including 1) following stress distribution, 2) connecting critical boundary regions, and 3) minimizing turning angles. Continuous spatial toolpaths are generated as iso-curves $\{\mathcal{T}\}$ of the field $P$ on each layer $\mathcal{S}$.


\subsection{Stress Field and PSL-guided Processing}\label{subsec:stressFieldProcessing}
Given a tetrahedral mesh and its boundary conditions, the stress tensor $\mathbf{T}(e)$ for each tetrahedral element $e\in\mathcal{M}$ is first computed. The principal stress $[ \sigma_1, \sigma_2, \sigma_3 ]$ is then obtained through singular value decomposition (SVD) of $\mathbf{T}(e)$, and the maximum principal stress $\sigma_{max} = \sigma_1$ and its corresponding principal stress direction $\Vec{\sigma}_{max}(e)$ can be identified by sorting the absolute values as $|\sigma_1| > |\sigma_2| > |\sigma_3|$. Akin to the objective of reinforcement proposed in our previous work~\cite{fang2020reinforced}, we aim to make the gradient of the scalar field $\nabla G(e)$ -- which dictates the growing direction of the curved printing layers -- perpendicular to the direction of $\Vec{\sigma}_{max}(e)$. We define an objective $E_{sf}(e)$ as $\nabla G(e) \cdot \Vec{\sigma}_{max}(e) = 0$. This ensures that the curved layers are generated aligning with the critical stress directions subsequently, which thereby reduces the chance of delamination between layers -- i.e., optimizes the fiber reinforcement.

\begin{wrapfigure}[6]{r}{0.4\linewidth}\vspace{-20pt}
\begin{center}
\hspace{1pt}\includegraphics[width=1.0\linewidth]{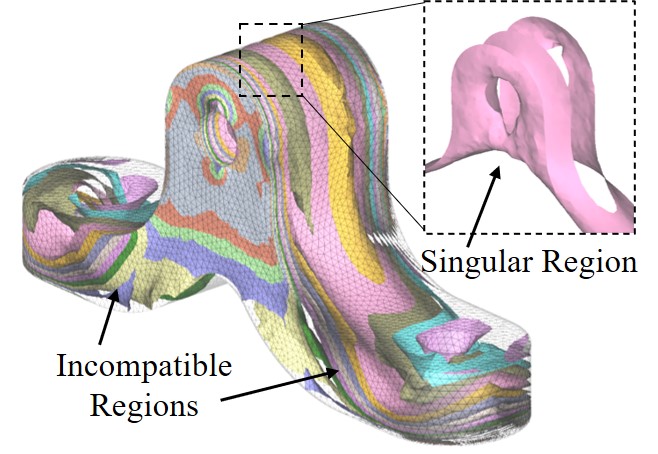}
\end{center}
\end{wrapfigure}
Directly minimizing $E_{sf}$ for all elements in the design domain can easily result in a non-harmonic scalar field. Additionally, it does not consider the different level of stresses on those critical elements that are more prone to causing material failure. As illustrated in Fig.~\ref{fig:framework}(b), the stress field $\mathcal{V}$ formed by $\{ \Vec{\sigma}_{max}(e) \}$ is a bidirectional vector field that presents incompatibilities in singular regions (e.g., bolted joints) where $\sigma_1\approx\sigma_2$.~\rev{}{An example of directly computing the guidance field and generating iso-surfaces without filtering out singularities and incompatible regions has been given in the warp figure. This leads to iso-surfaces that cannot be printed.} To avoid this, a field processing step is needed to filter the non-harmonic region and evaluate the importance of regions to formulate a more effective objective $E_{sf}$. In contrast to the approach suggested in~\cite{chen2022field} where the principal stress direction is subject to change by post-filtering, here we conduct a more effective strategy utilizing the principal stress line (PSL) to resolve issues of incompatibility and singularity in $\mathcal{V}$. The PSLs naturally adhere to the principal stress direction across the model while providing excellent references for aligning continuous fibers. Moreover, it aids in pinpointing the load-bearing regions in the model that necessitate a more stringent control of layer and fiber growth directions. This process takes into account the stress levels and respects the inherent properties of continuous fiber reinforcement.

We first introduce the algorithm below to trace PSLs (denoted by ${\mathcal{L}}$) as a streamline of the bidirectional vector field $\mathcal{V}$ (see also the illustration in Fig.~\ref{fig:framework}(c)). 
\begin{enumerate}
\item Starting from the center point $v_c(e_0)$ of a tetrahedral element $e_0 \in \mathcal{M}$ as the source of a PSL, We generate a ray along the direction of $\Vec{\sigma}_{max}(e_0)$ and compute the intersection point $v_{1}$ at a face $f_0 \in e_0$. Count the iteration number as $i=0$.
\item If the intersection face $f_i$ is not on the boundary of $\mathcal{M}$, find the neighboring element $e_{i+1}$ of $e_{i}$ with $f_i = e_i \cap e_{i+1}$.
\item 
Start from $v_i$ to generate ray $\vv{v_iv_e}$ along the direction of $\pm \Vec{\sigma}_{max}(e_{i+1})$ and ensure the vector $\vv{v_iv_e}$ points inward of the element (i.e., enable $\vv{v_iv_e} \cdot \vv {v_iv_c(e_{i+1})} > 0$) and find the next intersecting face $f_{i+1}$ and intersecting point $v_{i+1}$ along the ray $\vv{v_iv_e}$. 
\item Iteratively repeat steps 2 and 3 until reach an intersecting face as a boundary face or the total length of the traced PSL exceeds threshold\footnote{We choose $L_{max}$ as 100 times of the average length of mesh edges in our implementation by experiments.} (i.e., $\sum_i ||v_{i+1} - v_{i}|| > L_{max}$).
\end{enumerate}
Every element (i.e., $\forall e \in \mathcal{M}$) serves as a source element to trace a single PSL $\mathcal{L}(e)$. Each PSL traced from the starting element $e$ produces a set $\{ e_{\mathcal{L}} \}$, encompassing all elements it crosses. Only PSLs that connect to critical boundary regions are selected to remain by checking the condition that $\exists~e_j,e_k \in\{ e_{\mathcal{L}} \}$ with $e_j\in\partial\mathcal{M}_F$ and $e_k\in\partial\mathcal{M}_L$. A set $\{\mathcal{L}\}$ is formed by the selected PSLs -- see the illustration in Fig.~\ref{fig:framework}(c). This set $\{\mathcal{L}\}$ is used to guide the processing and filtering of the original principal stress field $\mathcal{V}$, which is presented below.

The selected PSLs in set $\{\mathcal{L}\}$ naturally follow the stress direction and establish geometric connections between load-bearing regions. We count the number of PSLs intersecting with each element $e$ as $N_{PSL}(e)$ and use it as the weighting to evaluate the significance of an element in reinforcement concerning maximal stresses. The elements containing any selected PSL (i.e., $N_{PSL}(e) > 0$) define the \textit{critical} regions that contribute to the continuous fiber-based reinforcement. Elements without any selected PSL crossing through them (i.e., $N_{PSL}(e) = 0$) are directly excluded when evaluating $E_{sp}$. By this method, we can successfully filter out those singular regions in $\mathcal{V}$ containing incompatible principal stresses, thus eliminating their influence on the computation of a harmonic field $G$ in subsequent steps.





It is worth mentioning that while the PSLs are computed as globally continuous and smooth lines, they do not contain information for manufacturing sequences. As a result, PSLs cannot be directly used as printed paths for continuous fiber. In the following steps of our computational pipeline, we follow the strategy of first decomposing the model into layers, and then generating toolpaths to ensure the design objectives and their manufacturability.

\subsection{Guidance Field Computing and Curved Layer Slicing}~\label{subsec:fieldComputing}
After obtaining a proper weighting to evaluate the importance of an element, the stress-following objective is well defined and not being influenced by singularities of stresses exhibited in $\mathcal{V}$. The scalar field $G$ used to guide curve layer slicing can then be computed by solving the following optimization problem.
\begin{align}
      \arg \min_{G} \ \ & \omega_{sf} \underbrace{\sum_{e\in\mathcal{M}} N_{PSL}(e) \  || \nabla G(e) \cdot \Vec{\sigma}_{max}(e) ||^2}_{\mathrm{Stress~Following}~E_{sf}(\mathcal{M})} + \\  & 
      \omega_{cg} \underbrace{ \sum_{ f\in \mathcal{M},~e_i \cap e_j = f} \|\nabla G(e_i) - \nabla G(e_j)\|_2^2}_{\mathrm{Compatibility~in~Gradient}~E_{cg}(\mathcal{M})}  +   \\ &
      \omega_{cp} \underbrace{ \sum_{v\in\{v_{cp}\}} || m G(v) - \sum_{v_k\in\mathcal{M}_{ROI}} G(v_k) ||^2 }_{\mathrm{Continuity~Protection}~E_{cp}(\mathcal{M})} \label{eq:layerOpt}.
\end{align}
\rev{}{The parameters $\omega_{sf}$, $\omega_{cg}$, and $\omega_{cp}$ are selected as the weight to balance three objectives.} \rev{}{$m$ in Eq.(\ref{eq:layerOpt}) indicates the total number of vertices contained by $\{v_{cp}\}$.} Note that the scalar field $G$ is represented as field values $G(v)$ that are defined on vertices $\forall v \in \mathcal{M}$. The field value inside a tetrahedral element is determined by the linear interpolation of the vertex values. The gradient of field value $G(e)$ in each tetrahedral element $e$ is a constant vector as the linear combination of field values on $e$'s vertices (details can be found in \cite{fang2020reinforced}).

Minimizing the first objective $E_{sf}$ ensures the alignment of curved layers with the principal stresses in critical regions to maximize the reinforcement performance of continuous fibers and reduce the possible layer delamination. The second term $E_{cg}$ of the objective function is used as a regularization term to \rev{generate harmonic fields}{optimize the compatibility of gradients between neighboring elements}. This avoids generating new singular regions, maintains smoothness of the curved layers, and prevents the creation of highly curved concave regions that are not printable. This energy term evaluates the variation of $\nabla G(e)$ between neighboring tetrahedral pairs $e_{i}$ and $e_{j}$ that share the same inner face $f$. Additionally, ensuring \rev{harmony}{compatibility} in the gradient also optimizes the uniformity of the layer height of the matrix material, preventing under-extrusion or over-extrusion in practical printing experiments that could lead to material failure. \rev{}{Compared with other ways to control the range of thickness (i.e., minimizing $\sum(|\nabla G(e) | - c)^2$) proposed in~\cite{fang2020reinforced} or directly optimizing the harmonic of the scalar field (i.e., letting $\nabla^2 G = 0$), optimizing $E_{cg}$ leads to a linear form that can be solved more efficiently.} Thanks to PSL-based field processing to filter out non-harmonic regions in the stress field $\mathcal{V}$, our experiments have found that minimizing $E_{cg}$ can \rev{effectively}{indirectly} control the harmonicity on resultant scalar fields. This avoids the time-consuming process of manual selection and iteration-based vector-to-scalar transfer schemes proposed in \cite{fang2020reinforced}. 

\begin{figure}[t]
 \centering 
\includegraphics[width=\linewidth]{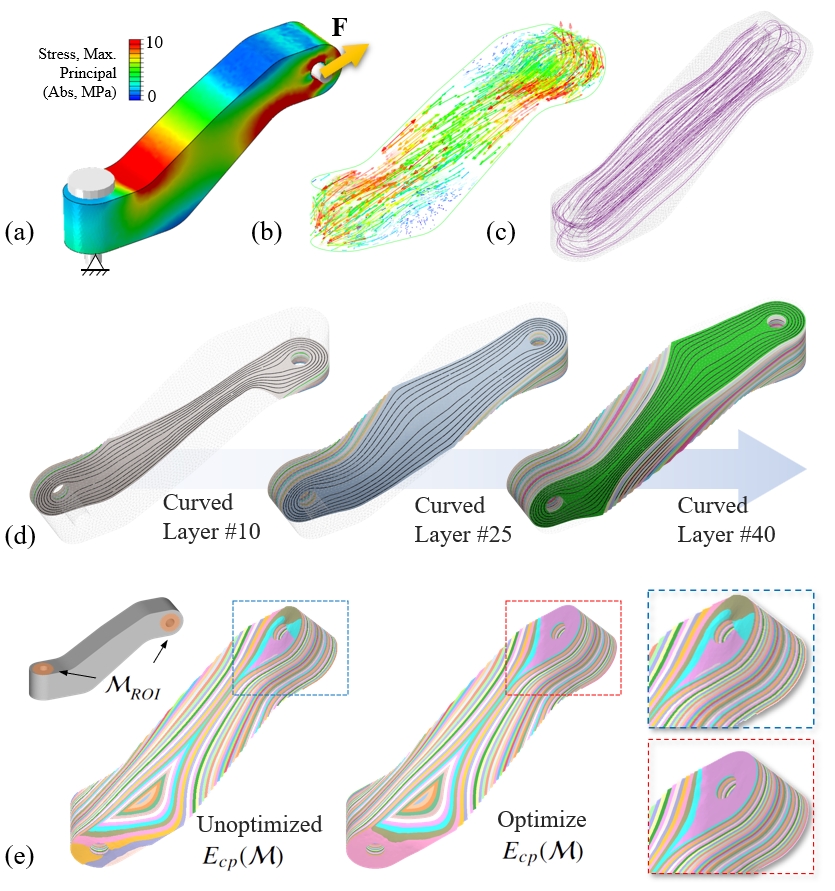}
\caption{(a-c) Stress field, principal stress directions, and traced principal stress lines for the Twist-Bar model under tensile loading. (d) Sequence of curved matrix layers and continuous fiber toolpaths aligned with the stress direction. (e) Without the continuity protection objective $E_{cp}(\mathcal{M})$, a layer with separated regions will be generated. After incorporating $E_{cp}(\mathcal{M})$ in optimization, curved layers that facilitate the requirement for connecting load-bearing regions can be generated.}
\label{fig:barModelResult}
\end{figure}

The third term, $E_{cp}$, is employed to preserve the geometric connectivity between the load-bearing regions, particularly those joints with bolt fixtures. Specifically, for each selected boundary region represented by the set $\{v_{cp}\}$ near $\partial \mathcal{M}_{F}$ or $\partial \mathcal{M}_{L}$, this function term enforces an equal scalar value to be assigned to the relevant vertices. This approach facilitates the requirement to generate larger surface layers in the user selected areas (denoted by $\mathcal{M}_{ROI}$). 
A comparison of applying $E_{cp}$ on bolt region to generate curved surfaces for fiber placement—with and without this term has been shown in Fig~\ref{fig:barModelResult}(c). Without applying $E_{cp}$, layers with separated regions are generated and toolpaths for fibers cannot effectively wind around the bolt regions. In our implementation, each region $\mathcal{M}_{ROI}$ to preserve fiber continuity is selected by users and set as 4-ring neighbors around $\partial \mathcal{M}_{F}$ and $\partial \mathcal{M}_{L}$ (see also the illustration in Fig.\ref{fig:barModelResult}(e)).

\begin{figure*}[t]
 \centering 
\includegraphics[width=\linewidth]{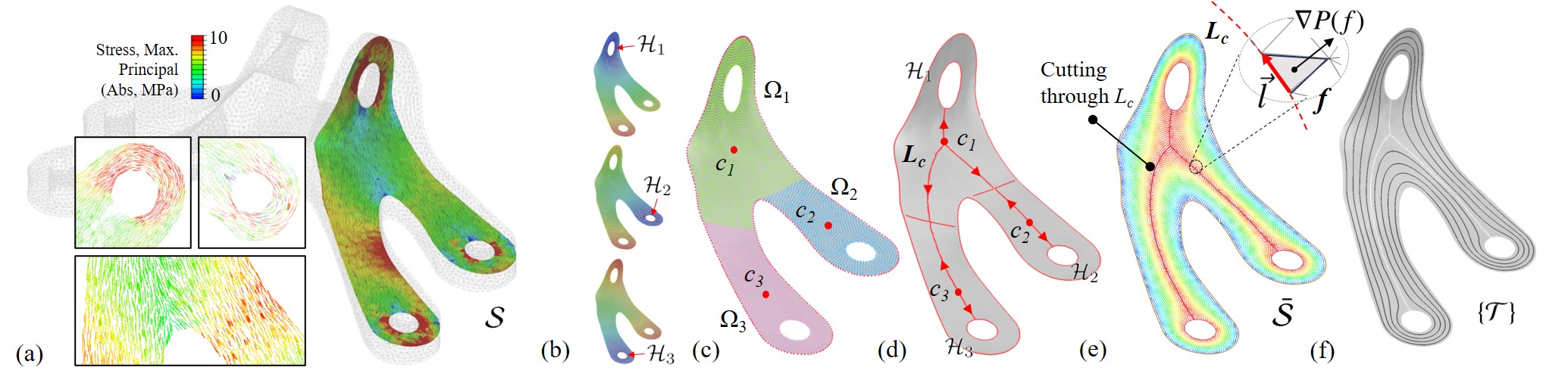}
\caption{Illustration of topology analysis based continuous \rev{CCF}{fiber} toolpath generation on a curved layer $\mathcal{S}$. 
(a) Stress flow is transferred onto the curved working surface. 
(b) A set of distance fields $\{ P_i \}$ is generated using the bolt joints (remarked as $\{ \mathcal{H}_i \}$) as the sources. 
(c) Voronoi tessellation is applied to the curved surface $\mathcal{S}$, categorizing the mesh into a set of regions $\{ \Omega_i \}$. The center-of-mass $\mathbf{c}_i$ of each region is computed using the strategy presented in~\cite{wang2015intrinsic}. 
(d) $L_c$ is established by connecting the shortest paths between $\mathbf{c}_i$ and the source region $\mathcal{H}_i$ (e.g., bolt joint on this model). 
(e) An updated mesh $\bar{\mathcal{S}}$ with all bolt joints being topologically connected is developed by cutting the original mesh with $L_c$. A guidance field $G$ is computed by minimizing the objectives listed in Eq.~\ref{eq:toolpathObjective}. 
(f) Continuous fiber toolpaths $\{\mathcal{T}\}$ are extracted as iso-curves of the field $P$.
}
\label{fig:toolpath}
\end{figure*}

\begin{figure}[t]
 \centering 
\includegraphics[width=\linewidth]{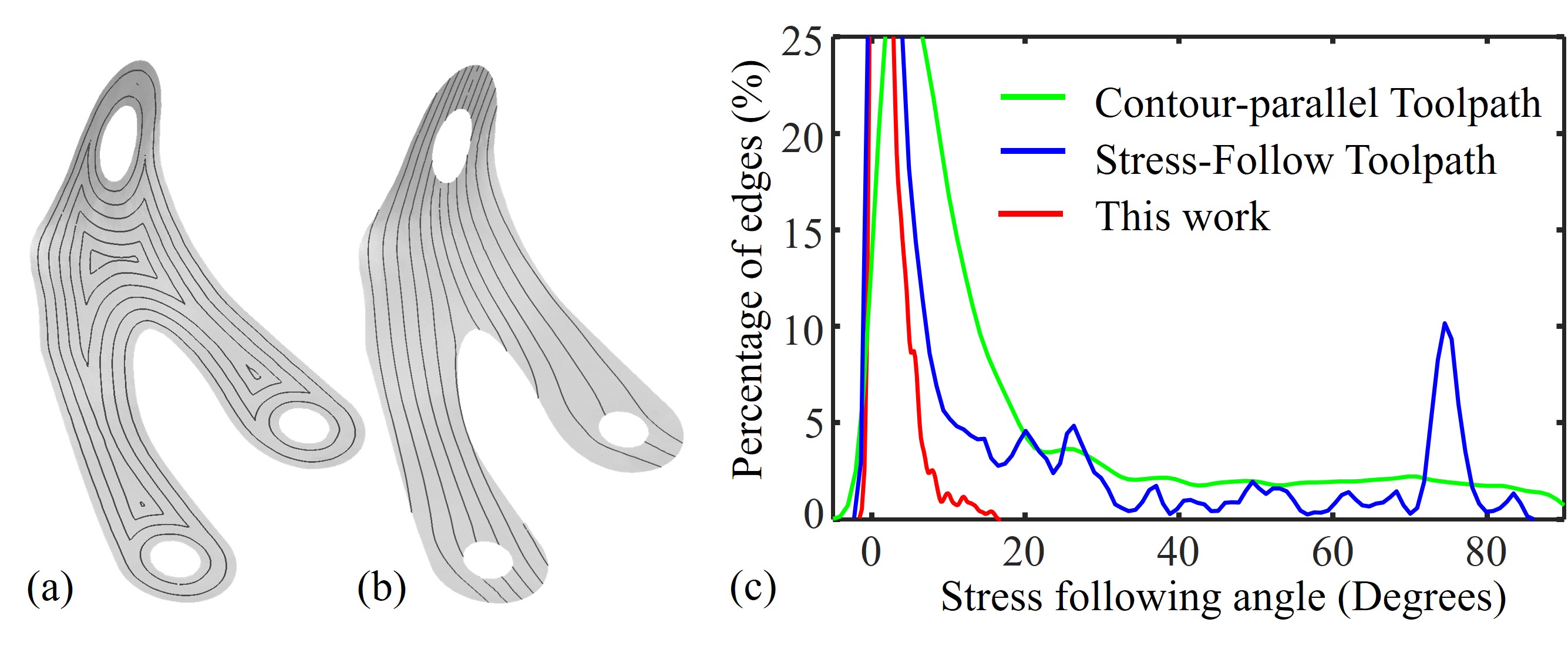}
\caption{(a) Contour-parallel toolpath generated as isocurves of the field using boundary nodes as sources. (b) Stress-follow toolpath presented in our previous work~\cite{fang2020reinforced, chen2022field}. (c) Statistical comparison of the stress following angles of different toolpaths, where the toolpath generated in this work can not only follow the stress distribution well but also remain continuous.}
\label{fig:toolpathComparsion}
\end{figure}
\begin{figure}[t]
 \centering 
\includegraphics[width=\linewidth]{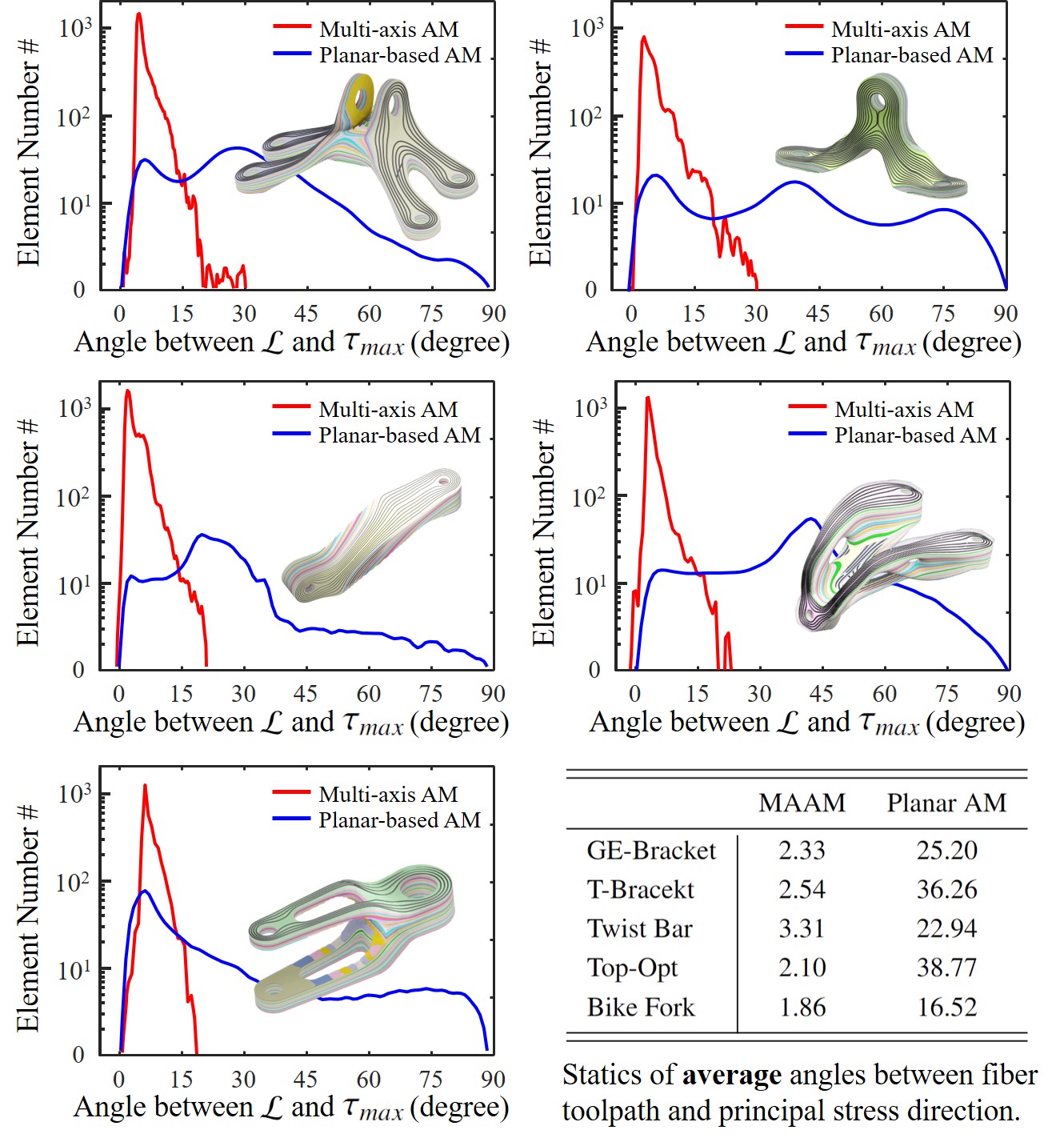}
\caption{Statistics to evaluate the level of alignment between the computed continuous fiber toolpaths $\mathcal{L}$ and the principal stress directions $\Vec{\sigma}_{\max}$ in the critical regions. The fiber toolpaths generated by our computational pipeline perfectly align with the principal stress direction, which lead to more effective reinforcement than those planar-based toolpaths for CCF printing.}
\label{fig:PSAlignAngle}
\end{figure}

This optimization problem can be effectively solved by computing the least-squares solution of 
a single linear system, which thereby is a highly efficient method. All examples presented in this paper can be computed within a few seconds -- the detailed computational statistics will be presented in Sec.~\ref{subsec:compResult}. Based on our experimental tests, $\omega_{sf} = 1.0$, $\omega_{cg} = 0.5$ and $\omega_{cp} = 0.1$ are used for all the examples \rev{}{(details will be discussed in Sec.~\ref{subsec:compResult})}. After computing a scalar field, the curved layers represented by a set of triangular meshes $\{\mathcal{S}\}$ are generated as iso-surfaces of $G$. 

\subsection{Continuous Fiber Toolpath Generation on Curved Layers}~\label{subsec:toolpathGeneration}
Given a curved layer generated by the aforementioned method, the spatial toolpaths for continuous fibers need to be computed on the layer by the objectives of reinforcement including the alignment with stress directions and the preservation of geometric continuity. The stress distribution on the curved layer, transferred from the original tetrahedral mesh $\mathcal{M}$, serves as the input of this process. Specifically, for each triangle face $f$ extracted from the tetrahedral element $e$ as part of an isosurface, we assign the principal stress value $\sigma_{max}(f) = \sigma_{max}(e)$. The principal stress direction $\Vec{\sigma}_{max}(f)$ on the face $f$ is computed by projection while taking into account the surface's normal as follows:
\begin{equation}~\label{eq:stressFieldProject}
    \Vec{\sigma}_{max}(f) = \Vec{\sigma}_{max}(e) - (\Vec{n}_f \cdot \Vec{\sigma}_{max}(e))~\Vec{n}_f.
\end{equation}
An example of the projected stress field onto a curved layer $\mathcal{S}$ is as visualized in Fig.~\ref{fig:toolpath}(a), where critical regions with high stress values can be easily identified and are displayed in red color. Note that the projection proposed in Eq.~\ref{eq:stressFieldProject} enables $\Vec{\sigma}_{max}(f)$ to align well with the surface without significantly altering the original stress direction (i.e., the difference between $\Vec{\sigma}_{max}(f)$ and $\Vec{\sigma}_{max}(e)$ are trivial in general). This nice result is mainly because of that our curved layer generation method already optimized $E_{sf}(\mathcal{M})$ to ensure $\Vec{\sigma}_{max}(e)$ locally aligns with the curved layer. Statistics for the angle between the final computed fiber toolpaths and the principal stress directions are given in Fig.~\ref{fig:PSAlignAngle} and presented in Sec.~\ref{subsec:compResult}. In summary, the angle between $\mathcal{L}$ and $\Vec{\sigma}_{max}(e)$ is controlled within 10 degrees in over 95\% regions.

After the stress distribution is projected onto the working surface, the objective of stress following similar to $E_{sf}$ can be adopted to generate stress-following toolpath. Direct optimizing stress-oriented objective on $\mathcal{S}$ is applied in our previous work~\cite{chen2022field, fang2020reinforced}. However, as illustrated in Fig.~\ref{fig:toolpathComparsion}(b), toolpaths generated in this way will bring discontinuity with many sharp turning angles. Additionally, this toolpath pattern does not guarantee stress-following at the bolt joint region where a circular-like direction is detected. A key factor is that the curved working surface contains different topology, and critical regions are disconnected. In this study, we tackled this issue by first applying topology analysis and modification to ensure that both the objectives of stress-following and continuity are met. The detailed process is described below with the help of illustration given Fig.~\ref{fig:toolpath}.
\begin{enumerate}
\item Within user-defined critical regions (e.g., bolt joints), detect a set of connected boundary contours $\{\mathcal{H}_{i=1,2, ...}\}$.
\item Compute the geodesic distance field $P_i$ using vertices $v \in \mathcal{H}_i$ as sources.
\item Perform Voronoi tessellation to categorize $\mathcal{S}$ into a set of regions $\{\Omega_i\}$. For each vertex $v$ in $\Omega_i$, the Voronoi tessellation ensures that $P_i(v)$ is greater than any other field value (i.e., $P_i(v) = \max \{ P_1(v), P_2(v), ...\}$).
\item Compute the center of mass (CoM) point $\mathbf{c}_i$ on the curved surface for each region $\Omega_i$.
\item Compute the shortest path between CoM points and the boundary regions (e.g., the connecting path between $\mathbf{c}_1$ and $\mathbf{c}_2$, and the connecting path between $\mathbf{c}_1$ and $\mathcal{H}_{1}$). Connect all the paths to form the collection of curves as $L_c$.
\item Generate updated curved layer $\Bar{\mathcal{S}}$ by cutting the original mesh with $L_c$. This ensures that the critical regions $\{ \mathcal{H}_i \}$ are topologically connected.
\end{enumerate}

\rev{}{In \textit{Step 2}, we compute the geodesic distance field using the heat method presented in~\cite{crane2017heat}. This is followed by a Voronoi tessellation, which is executed by utilizing the greatest value of the distance field on each vertex. In \textit{Step 5}, the shortest paths are identified using a flooding algorithm that traverses from the source point until reaches the boundary of each region. After completing the process of topology analysis,} the scalar field $P(\cdot)$ for toolpath generation on the curved layer $\Bar{\mathcal{S}}$ is then computed by solving the following optimization problem:
\begin{align}
     &\arg \min_{P} \ \  
    \underbrace{\omega_{sf} \sum_{f~\in~\Bar{\mathcal{S}}} \ \sigma(f) / \sigma_{max} \cdot|| \nabla P(f) \cdot \Vec{\sigma}_{max} (f) ||^2}_{\mathrm{Stress~Following}~\mathcal{O}_{sf}(\mathcal{S})} ~+ \\ ~\label{eq:toolpathObjective}
    &\underbrace{\omega_{cp} \sum_{ l~\in~L_c} || \nabla P(f) \cdot \Vec{l}~||^2}_{\mathrm{Continuity~Protection}~\mathcal{O}_{cp}(\mathcal{S})} + ~\underbrace{ \omega_{hf}\sum_{l~\in~\Bar{\mathcal{S}},~f_i~\cap~f_j = l}||\nabla P(f_i) - \nabla P(f_j)||_2^2}_{\mathrm{Compatibility~in~Gradient}~\mathcal{O}_{cg}(\mathcal{S})}.
\end{align}
Again, the field $P(\cdot)$ is represented by field values defined on vertices, which are the unknown variables to be determined. The field value inside a triangle is determined by the linear interpolation of the vertex values.

The first objective $\mathcal{O}_{sf}$ requires the gradient of the scalar field $P$ is perpendicular to $\Vec{\sigma}_{max}$. This thereafter makes the computed toolpath, which will be extracted from the iso-curves of $P$, well aligned with the stress directions. 
The second objective, $\mathcal{O}_{cp}$, is only applied to the triangles connected to $L_c$. This ensures the geometric continuity of resultant fiber toolpaths, especially in load-bearing regions such as bolt joints. As these regions are topologically connected in the surface patch $\Bar{\mathcal{S}}$ processed by $L_c$, aligning the field gradient with the local direction of $L_c$ allows the fiber toolpath to naturally wind around all critical regions $\mathcal{H}$. Here the direction of $L_c$ is locally defined by $\Vec{l}$, as shown in the zoom-in view of Fig.~\ref{fig:toolpath}(e).
%
The third objective $\mathcal{O}_{cg}$ is to \rev{generate a harmonic field}{ensure the compatibility of the gradients, which is applied on adjacent triangle faces (i.e., $f_i$ and $f_j$) that share a common edge $l$}. This enhances the smoothness of the toolpaths and prevents sharp corners that could potentially damage the fibers during the fabrication process. An example of fiber breakage caused by sharp turns~\cite{huang2023turning} is indicated by a dotted red circle in Fig.~\ref{fig:topoptResult}(b5, b6).

\begin{figure}[t]
 \centering 
\includegraphics[width=\linewidth]{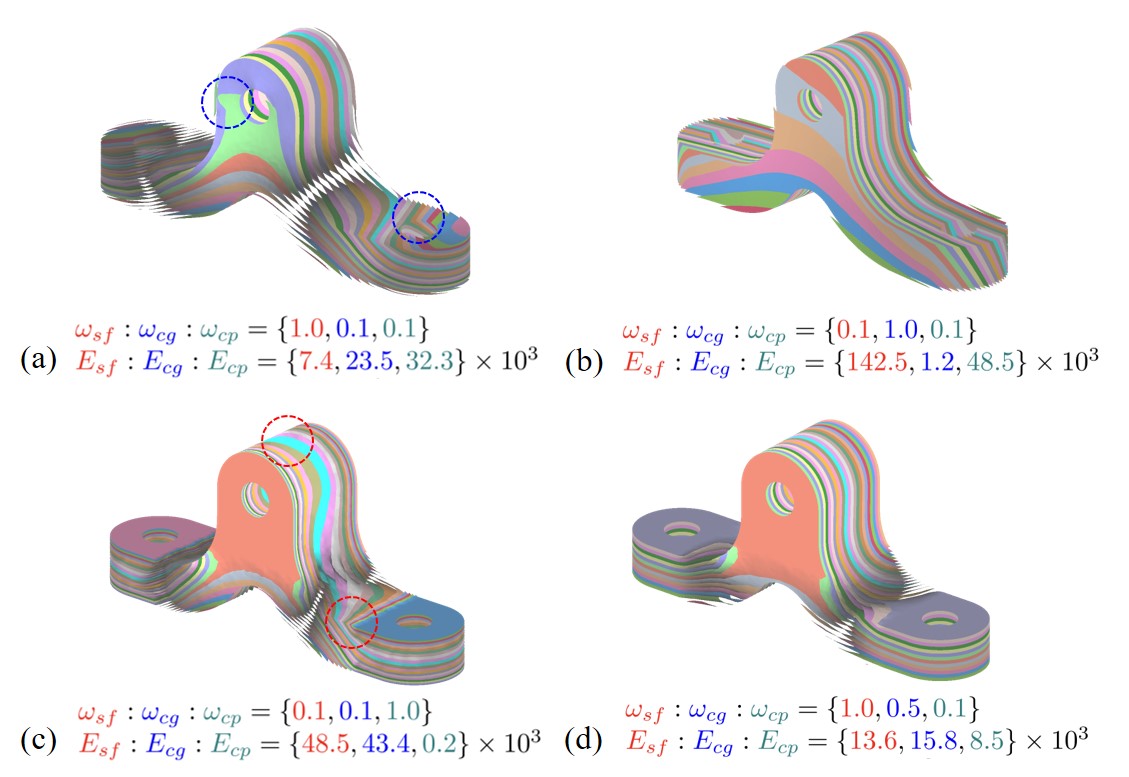}
\caption{\rev{}{The study of using different weights to balance different objects and their impact on the resultant curved layers for printing.
Residuals of all energy terms are given. 
}
}
\label{fig:weightCompare}
\end{figure}

\begin{figure*}[t]
 \centering 
\includegraphics[width=\linewidth]{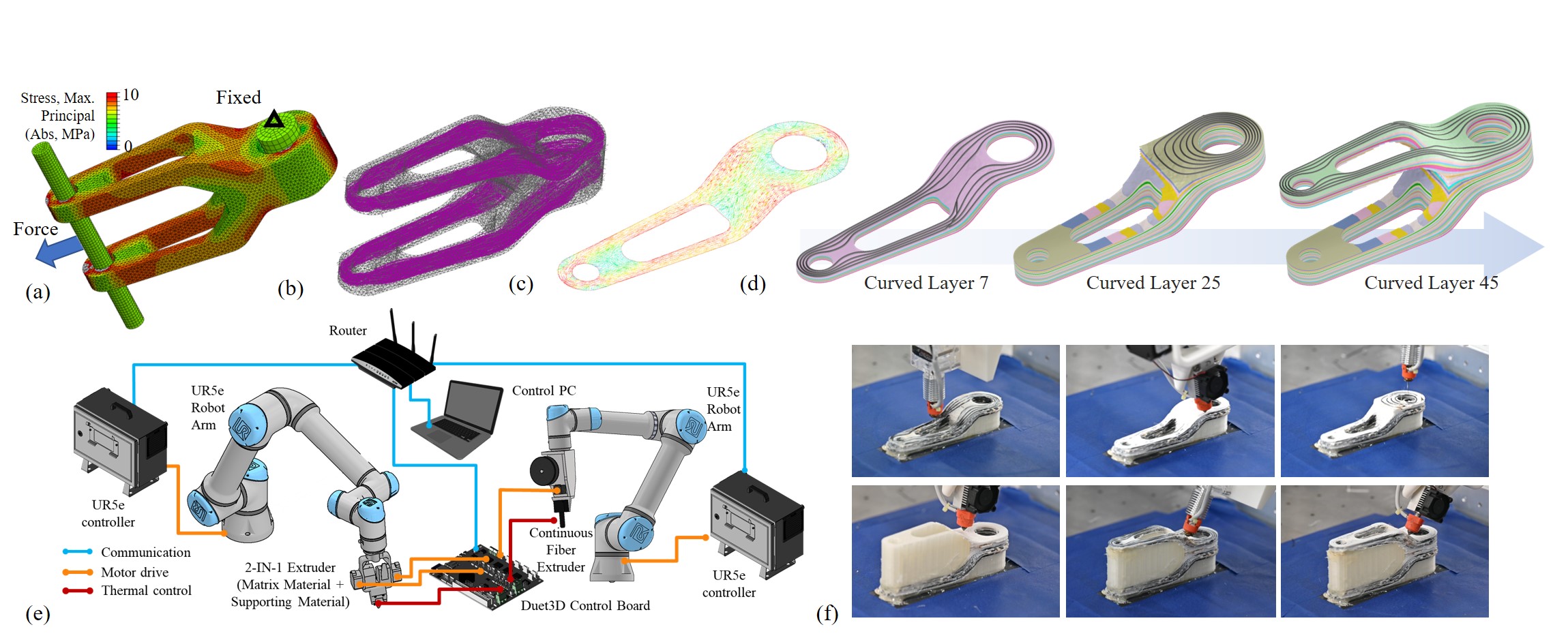}
\caption{Computational results and the MAAM fabrication process of the Bike Fork model. (a) Stress distribution and (b) traced PSLs after filtering. (c) Stress distribution projected onto a curved layer. (d) Sequence of computed curved layers for matrix material and corresponding spatially-placed continuous fiber toolpaths on each layer. (e) Hardware illustration of the MAAM system with dual robotic arms. (f) Fabrication process of the Bike Fork model, demonstrating the switching between supporting structure, thermoplastic matrix, and continuous fiber composites. \rev{Note that the layers for supporting structures are generated by the method presented in~\cite{Zhang23ICRA}.}{}
}
\label{fig:forkResult}
\end{figure*}

\begin{figure}[t]
 \centering 
\includegraphics[width=\linewidth]{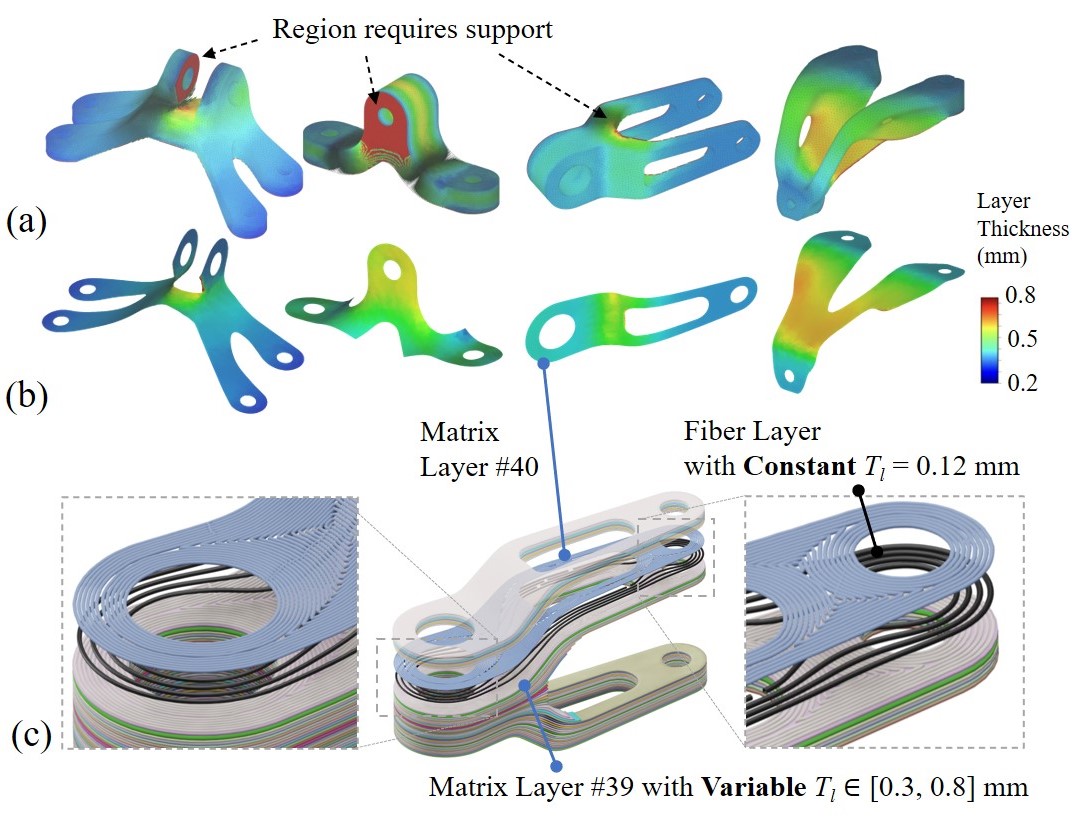}
\caption{\rev{}{(a) Illustration depicting the layer thickness of the matrix material -- visualized by colors. (b) Distribution of layer thickness across a few working surfaces that are important for fiber reinforcement. (c) Sequence of layers and toolpaths for the layers of matrix material and continuous fibers.}
}
\label{fig:layerThickness}
\end{figure}

\begin{figure*}[t]\centering 
\includegraphics[width=\linewidth]{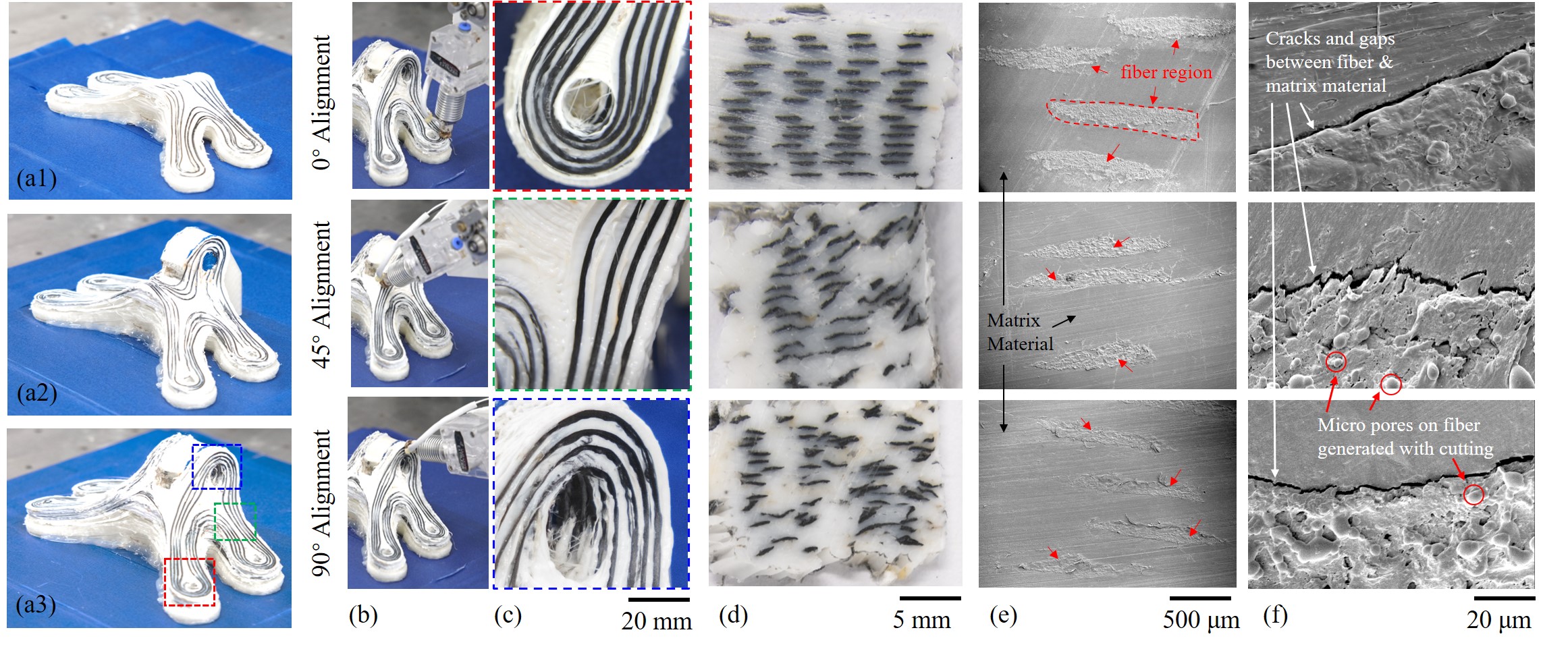}
\caption{Fabrication verification of our CFRTPCs toolpaths in 3D space for the GE-bracket model. (a1-a3) Printing results for layers 14, 22, and 40, which follow the computational results very well. (b) Demonstration of fiber alignment in different \rev{titling}{tilting} angles. (c) zoom-in views reveal that the continuous fiber are perfectly attached to the matrix material. (d) Images of the cross-sections show that printing along our 3D spatial toolpaths can build impeccable bonding for different fiber alignment angles. (e,f) scanning electron microscope (SEM) images display the interface between the matrix and the fibers.\rev{, where the gap is controlled to within $3\mu m$.}{}}
\label{fig:GEBFabResult}
\end{figure*}

\begin{figure*}[t]
 \centering 
\includegraphics[width=0.8\linewidth]{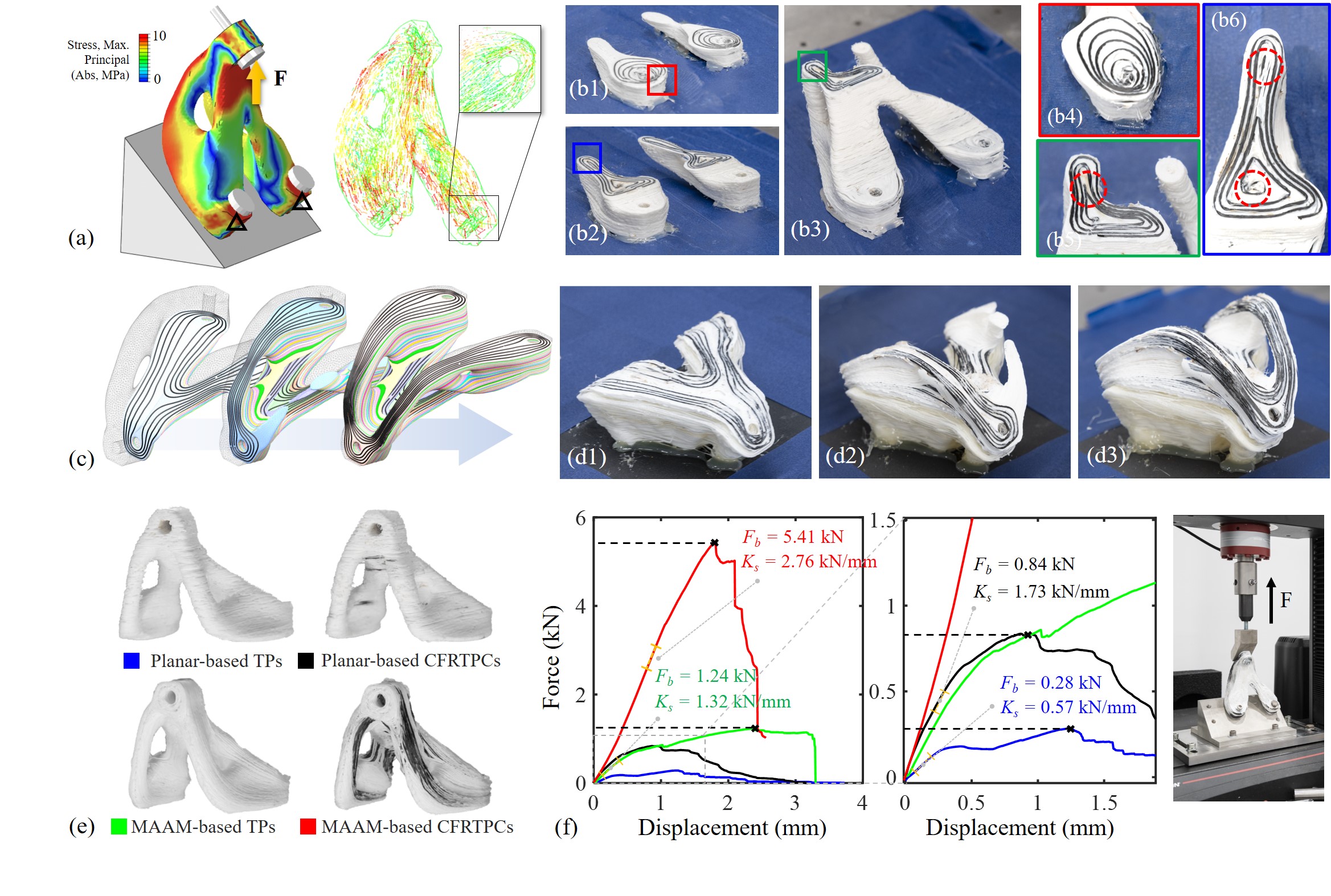}
\vspace{-15pt}
\caption{(a) Stress distribution of the Top-Opt model under loading. (b1-b6) Fabrication process of the model using planar-based toolpaths. (c) Curved layers of the model optimized according to the stress distribution. (d1-d3) Spatial printing of continuous fibers by MAAM using our toolpaths, which guarantee the stress adherence and the continuity for enhanced reinforcement. (e) Resultant models fabricated by different strategies. (f) Tensile test validation on physical models, demonstrating the exceptional mechanical performance of MAAM with CFRTPCs using our toolpaths.}
\label{fig:topoptResult}
\end{figure*}

\section{\rev{Result and Discussion}{Details of Implementation and Results}}~\label{Sec:Result}
In this section, we first present the computational results of our pipeline. After that, we demonstrate successful fabrication of mechanical components with complex geometry and stress distribution by using a system with dual robotic arms. The tensile tests verify the high effectiveness of our method in generating spatial fiber toolpaths for mechanical strength reinforcement. All results of curved layers, toolpaths, the fabrication process, and the tensile tests can also be found in the supplementary video at:~\url{https://youtu.be/7Jxyu9uRMLo}.


\subsection{\rev{Results of Computation}{Computational Details and Results}}\label{subsec:compResult}
The computational framework is implemented in C++ and was tested on a laptop equipped with an Intel i7-12700H CPU (24 Core, 3.50 GHz) and 32 GB of RAM. The Eigen library~\cite{Eigen} is used to solve large linear systems. The robustness of our method has been tested by a variety of tests on components with freeform shapes under different loads. \rev{}{For all examples, we utilize the  software Abaqus to perform FEA with standard/explicit model, and Tetgen~\cite{hang2015tetgen} is employed to generate isotropic tetrahedral meshes. The tetrahedral mesh is incorporated into Abaqus as C3D10 elements, and we have included connecting rods and screw fixtures in a contact-aware model to reflect the stress direction under loading conditions more precisely.}

\begin{table*}[t]
\centering 
\caption{Statistics of our computational pipeline for generating curved layers and spatial toolpaths generation}\label{tab:CompStatistic}
\footnotesize \vspace{5pt}
\begin{tabular}{r|c|r||r|c|c||c|c|c|c|c||r}
\hline 
        &   &   &  \multicolumn{3}{c||}{Comp. Time for Curved Layers (sec.)}  &   &  \multicolumn{4}{c||}{Avg. Comp. Time (sec.) for Toolpaths per Layer$^\dag$}   &   Total \\ 
\cline{4-6}  \cline{8-11}
{Model} & {Fig.} &  Tet~\#  & {PSL} & {Opt. Field $G$} & Extract $\mathcal{S}$ &  Layer \# & Proj. $\tau_{max}$ &  Seg.$^{*}$  & Comp. $P$ & Extract $\mathcal{L}$ &  {Time (sec.)} \\ 
\hline\hline \specialrule{0em}{1pt}{1pt}
GE-Bracket &  \ref{fig:teaser}, \ref{fig:toolpath}  &  172,169 &  17.64 &  3.81 & 2.24 &  60 &  0.72 &  5.03 & 8.21 & 1.69 & 962.69  
\\ \specialrule{0em}{1pt}{1pt}
T-Bracket &  \ref{fig:framework}, \ref{fig:TBracketResult} & 85,119  &  7.68 &  1.26 & 1.78 &  45 &  0.34 &  3.44 & 6.46 & 1.22 & 187.54  \\ \specialrule{0em}{1pt}{1pt}
Twist Bar &  \ref{fig:barModelResult} & 82,770  &  9.85 &  3.29 & 1.92 & 50 &  0.19 &  1.31 &  4.28 & 0.93 & 339.00  
\\ \specialrule{0em}{1pt}{1pt}
Bike Fork & \ref{fig:forkResult}, \ref{fig:forkFabrication}  & 107,097  &  14.11 &  3.12 & 1.41 & 50 &  0.45 &  6.23 & 4.57 & 4.35 & 797.50  
\\ \specialrule{0em}{1pt}{1pt} 
Top-Opt &  \ref{fig:topoptResult} &  70,505 &  11.43 &  2.50 & 2.24 & 60 &  0.77 &  4.23 & 8.37 & 1.79 & 1,022.40  
\\ 
\hline
\end{tabular}
\vspace{-5pt}
\begin{flushleft}
~~~~$^\dag$~This includes computing time for generating toolpaths for both the matrix material and the continuous fiber material on a curved layer. \\
~~~~$^*$~This includes geometry-based center line detection (as illustrated in Fig.~\ref{fig:toolpath}(c)) and layer segmentation.
\end{flushleft}
\end{table*}

The first model tested is the GE-Bracket component, where the computational result can be seen in Figs.~\ref{fig:teaser} and~\ref{fig:toolpath}. This model has four bottom holes fixed on a plane using screws, and tensile loading was applied to the two top through holes. It can be found from Fig.~\ref{fig:teaser}(a) that the principal stress direction of the model changes dramatically and the stress flow connects all the bolt regions. As observed in Fig. \ref{fig:teaser}(d), the computed curved layers align well with the principal stress directions. Various patterns of continuous fiber toolpath are successfully computed on layers (see the result of layer 14, layer 22, and layer 40 in Fig.~\ref{fig:teaser}(d)), and all of them ensure fully continuous \rev{CCF}{fiber} toolpaths. The computed spatial fiber toolpath optimizes the turning angle and perfectly follows the stress directions (especially at the hole region). The sequence for material accumulation along stress flow directions is guided by the computed toolpaths, therefore maximizing reinforcement by capitalizing on the anisotropic properties of the continuous fibers. 


The static analysis of toolpath alignment with the principal stress has been presented in Fig.~\ref{fig:PSAlignAngle}. The angles between toolpaths and the maximal stresses are measured in the critical regions as determined by our method presented in Sec.~\ref{subsec:stressFieldProcessing}. It can be seen that the average angle decreased from $25.2^{\circ}$ (planar-layer based AM) to $2.33^{\circ}$ using our toolpaths for MAAM. Consistently positive results can be observed in four other components tested in our work, as shown in Fig.~\ref{fig:PSAlignAngle}, including the T-Bracket (Figs.~\ref{fig:framework} and \ref{fig:TBracketResult}), the Twist-Bar model (Fig.~\ref{fig:barModelResult}), the Bike-Fork model (Figs.~\ref{fig:forkResult} and \ref{fig:forkFabrication}) and the Top-Opt model~\cite{fang2020reinforced} (Fig.~\ref{fig:topoptResult}). In all these tests, the final computed continuous fiber toolpath perfectly follows the stress direction in the critical region, with an average angle of alignment less than $3.31^{\circ}$.

\rev{}{When optimizing the scalar field $G$, we solve the problem in a least squares form and try to balance among various objectives by changing the weight of each energy term. The influence of weights $\omega_{sf}$, $\omega_{cg}$, and $\omega_{cp}$ on the residuals and the resultant curved layers of the T-Bracket model has been studied and illustrated in Fig.\ref{fig:weightCompare}. When the weight $\omega_{sf}$ is significantly larger than the other two weights, the resultant iso-surfaces strictly follow the stress direction (with the lowest residuals in $E_{sf}$) but fail to maintain connectivity. Small patches are generated and highlighted by blue circles in Fig.\ref{fig:weightCompare}(a), which is similar to the example discussed in Fig.\ref{fig:barModelResult}(e). Conversely, choosing a weight $\omega_{cg}$ for compatibility-in-gradient much larger than the other two leads to iso-surfaces that are almost parallel with each other (as shown in Fig.~\ref{fig:weightCompare}(b)), indicating less consideration for fiber reinforcement. If the weight for continuity protection $\omega_{cp}$ is set as much higher, it will generate incompatible regions near the boundary (highlighted by red circles in Fig.~\ref{fig:weightCompare}(c)), failing to control layer thickness for curved layers. After conducting numerical tests, we selected the weights as shown in Fig.~\ref{fig:weightCompare}(d), which can achieve the lowest residuals and an effective balance among different objectives.}

Our computational pipeline also shows high efficiency -- see the comprehensive computational statistics listed in Table~\ref{tab:CompStatistic}. The efficiency primarily stems from formulating the problems of toolpath and curved layer generation as convex optimization of field-based computing. As a result, all the computation for complex models encompassing more than $100k$ tetrahedra can be completed in less than 20 minutes. \rev{}{The computational statistics reported in Table~\ref{tab:CompStatistic} are based on selecting a mesh density with an average edge length $l_{e}$ as $2.5\%$ of the model size. This choice well balances the numerical accuracy with the computational cost. By using this mesh density,} the computing time for all models is significantly less than the time required for fabrication. \rev{}{Further discussion on the impact of mesh density can be found in Sec.~\ref{subsec:discussion}.}

\begin{table*}
\caption{Fabrication statistics and tensile tests on models fabricated by planar-based AM and MAAM.}
\vspace{5pt}
\centering\label{tab:comparisonangleave}
\footnotesize
\begin{tabular}{r|r||c|c|c||c|c||c|c}
\hline
\specialrule{0em}{1pt}{1pt}
 & & Dimensions & \multicolumn{2}{c||}{Material Usage$^*$: Matrix (g) + Fiber (mm)}  & \multicolumn{2}{c||}{Failure load $F_b$ (kN)} & \multicolumn{2}{c}{Model Stiffness $\sigma_s$ (kN/mm)}\\
\specialrule{0em}{1pt}{1pt}
\cline{4-9} 
\specialrule{0em}{1pt}{1pt}
Models & Fig. & ($mm \times mm \times mm$) & Planar-based AM & Multi-Axis AM & Planar AM & Multi-Axis AM & Planar AM & Multi-Axis AM\\
\specialrule{0em}{1pt}{1pt} \hline\hline \specialrule{0em}{1pt}{1pt}
GE-Bracket   & \ref{fig:GEBFabResult} &  $174 \times 105 \times 62 $ &  $203.7g + 17.5m$  &  $197.7g + 18.1m$ &  $2.35$   & 
$12.32~(\uparrow424.2\%)$ &  $1.81$ &  $3.12~(\uparrow72.4\%)$  \\ 
\specialrule{0em}{1pt}{1pt}
Top-Opt  & \ref{fig:topoptResult} &  $108 \times 98 \times 123$ &  $142.8g + 13.3 m$  &  $161.3g + 14.7 m$ &  $0.84$   & 
$5.41~(\uparrow544.0\%)$ & $1.73$ &  $2.76~(\uparrow59.5\%)$ \\ 
\specialrule{0em}{1pt}{1pt}
Bike Fork & \ref{fig:forkFabrication} &  $143 \times 42 \times 60$ &  $100.8g + 5.4m$  & $96.7g + 6.0m$  &  $3.46$   & $9.27~(\uparrow167.9\%)$ &
$0.86$ &  $1.49~(\uparrow73.2\%)$  \\ 
\specialrule{0em}{1pt}{1pt}
T-Bracket & \ref{fig:TBracketResult} &   $130 \times 30 \times 65$ & $60.2g+7.3m$  &  $63.5g+7.1m$ &  $~~3.16^\dag$ & 
$6.48~(\uparrow105.1\%)$ &  $~~0.82^\dag$ & $~~1.97~(\uparrow140.2\%)$ \\ 
\specialrule{0em}{1pt}{1pt}
\hline
\end{tabular}
\vspace{-5pt}
\begin{flushleft}
$^\dag$~For the T-Bracket model, the strongest result for planar-based AM was selected (i.e., planar layers sliced along the x-axis) -- as illustrated in Fig.~\ref{fig:TBracketResult}. \\
$^*$~\rev{}{Weighting of matrix material and the length of continuous fiber are measured from the physical fabrication setup.}
\end{flushleft}
\end{table*}    

\subsection{Robot-Assisted MAAM Platform}~\label{subsec:hardware}
To realize spatial alignment of continuous fiber, we built a robot-assisted multi-axis additive manufacturing hardware setup. We utilized the out-of-nozzle impregnation strategy for fabrication, where the matrix material and the continuous fibers are aligned sequentially. To facilitate multi-axis motions, two 6-DoFs UR5e robot arms with the repeatability precision at 0.05 mm are employed (as shown in Figs.~\ref{fig:teaser}(c) and \ref{fig:forkResult}(e)). The left robot arm is equipped with a duo-extruder printing head designed in our previous work~\cite{Zhang23ICRA}, which facilitates the extrusion of both thermoplastic and supporting materials (e.g., polyvinyl alcohol). Conversely, the right robot arm houses a printing head equipped with an individual continuous fiber printer head that was independently constructed. The materials chosen for fabricating the models are Euson 1.75 mm PLA and Markforged CF-FR-50 (with cross-section diameter as 0.37 mm) for matrix and fiber materials respectively. The nozzle diameter is selected as $0.8~mm$ for the matrix material, \rev{and the layer height is maintained at around $0.6~mm$ with}{which is capable of dynamically changing the layer height by adjusting the extrusion rate of the nozzle~\cite{etienne2019curvislicer,zhang2022s3}. The variation in layer height for the curved layers on matrix material, demonstrated in Fig.~\ref{fig:layerThickness}(a), is determined by measuring the distance from each waypoint $p\in\mathcal{T}_i$ generated from layer $\mathcal{S}_i$ to the previous layer $\mathcal{S}_{i-1}$. By optimizing the compatibility in the gradient field (i.e., minimizing $E_{cg}$), we can control the thickness in most regions within the range of $[0.3, 0.7]~\mathrm{mm}$.} The printing temperature of matrix material is set at $210^{\circ}\mathrm{C}$. On the other hand, the continuous fiber printing head utilizes a nozzle with rounded corners to better assist in compressing the fiber filament onto the resin matrix, maintaining a \rev{}{constant} layer height of 0.12 mm with the printing temperature at $250^{\circ}\mathrm{C}$.

In our robot-assisted MAAM system, we employ a Duet3D control board integrated with the Marlin control framework to synchronize robot motion with material extrusion. Based on the toolpath $\mathcal{L}$ that is represented by a set of waypoints with both position and orientation, the RoboDK software is utilized to plan the smooth and continuous motion of the robot arm in the configuration space. During the fabrication process, we maintain the line speed of the end-effector constantly at 25 mm/min and 8 mm/min for the thermoplastic and continuous fiber materials respectively. This ensures a tight bond between materials, \rev{}{and the printing sequence has been shown in Fig.~\ref{fig:layerThickness}(c). In local regions where the matrix material overlaps with fibers printed in previous layers, we reduce the extrusion rate of material to avoid over accumulation.} The processes of switching between different nozzles during the fabrication of the bike-fork model are illustrated in Fig.~\ref{fig:forkResult}(e). 

\subsection{Results of Fabrication}~\label{subsec:fabricationRes}
Fabrication results of the GE-Bracket model have been verified as shown in Fig.~\ref{fig:GEBFabResult}. It can be observed that the continuous fibers are aligned in 3D along different orientations from vertical to horizontal (i.e., $[0^\circ,90^\circ]$), all achieving perfect and tight alignment with the matrix material. Based on the optimized toolpaths generated by our approach, the MAAM system can realize precise control of the position and orientation for fiber alignment. \rev{The effectiveness of the fabrication process is further verified through zoom-in views shown in Fig.~\ref{fig:GEBFabResult}(c) as well as from}{To verify the bonding between matrix and fiber material, we used a cutting machine (ATM - Brilliant 220) with a diamond wheel (LECO Instruments (UK) LTD) to cut the printed component and examine the cross-section of the CFRPTCs, with results shown in Fig.~\ref{fig:GEBFabResult}(d). During the cutting process, cooling water was applied to bind the dust produced from sawing and wash away chips. After cutting, the surface was cleaned with ethanol. From the captured } scanning electron microscope (SEM) images, \rev{of the cross-section in the corresponding regions.
As shown in Fig.~\ref{fig:GEBFabResult}(f), the gap}{we could observe that the width of cracks} between the fiber and matrix material is controlled to be less than $3~\mu m$. This demonstrated the effectiveness of MAAM in fabricating CFRTPCs with tight \rev{bounding}{bonding} between the fiber and the matrix materials.

The comparison of fabrication results by using planar-based layers and our spatial toolpaths has been given in Fig.~\ref{fig:topoptResult}. It can be seen that applying planar layers to slice the model fails to align the fibers with the principal stress directions. More seriously, it generates a group of small patches where fibers cannot be placed. Moreover, the sliced layer contains narrow and separate regions that only offer limited space for fiber placement. In particular, sharp corners in the contour-parallel toolpath cannot be avoided, which leads to fiber breakage during the printing process and has been reported in~\cite{huang2023turning} -- see also those regions highlighted by red circles in Fig.~\ref{fig:topoptResult}(b6). All those factors will cause material failure under loading. More will be demonstrated via tensile results presented in the following sub-section. In contrast, by employing the optimized spatial toolpaths and MAAM strategy, fibers can align successfully on curved matrix layers while forming a continuous pattern along the principal stress directions to facilitate effective reinforcement. The fabrication results can be found in Fig.~\ref{fig:topoptResult}(d1-d3)) and also in the supplementary video. \rev{}{For all examples shown in this paper, the curved supporting structures were generated using the method proposed in our previous work~\cite{Zhang23ICRA}}. The water-soluble PVA material is used for supporting structures in all fabrication experiments.

\begin{figure}[t]
 \centering 
\includegraphics[width=\linewidth]{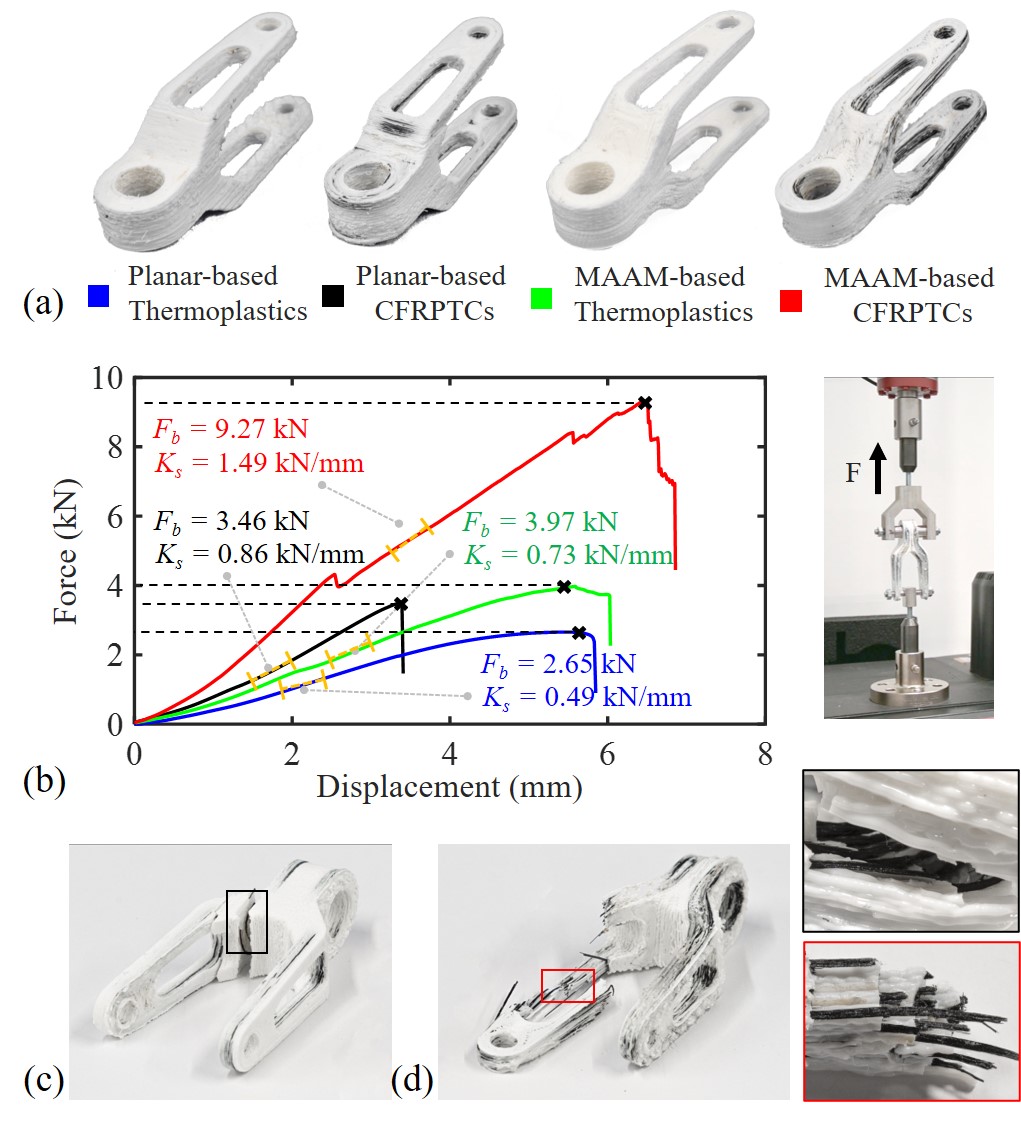}
\caption{Physical experimental results for the Bike Fork model. (a) Results of the specimens fabricated by different strategies. (b) Tensile test results on the specimens. (c) Breakage of the planar-based CFRTPC model occurred due to layer delamination, as the fibers do not align with the stress flow.
(d) Differently, fiber breakage is demonstrated on the model fabricated by MAAM using our toolpaths, which underscores the high efficacy of fiber reinforcement.}
\label{fig:forkFabrication}
\end{figure}

\begin{figure*}[t]
 \centering 
\includegraphics[width=0.8\linewidth]{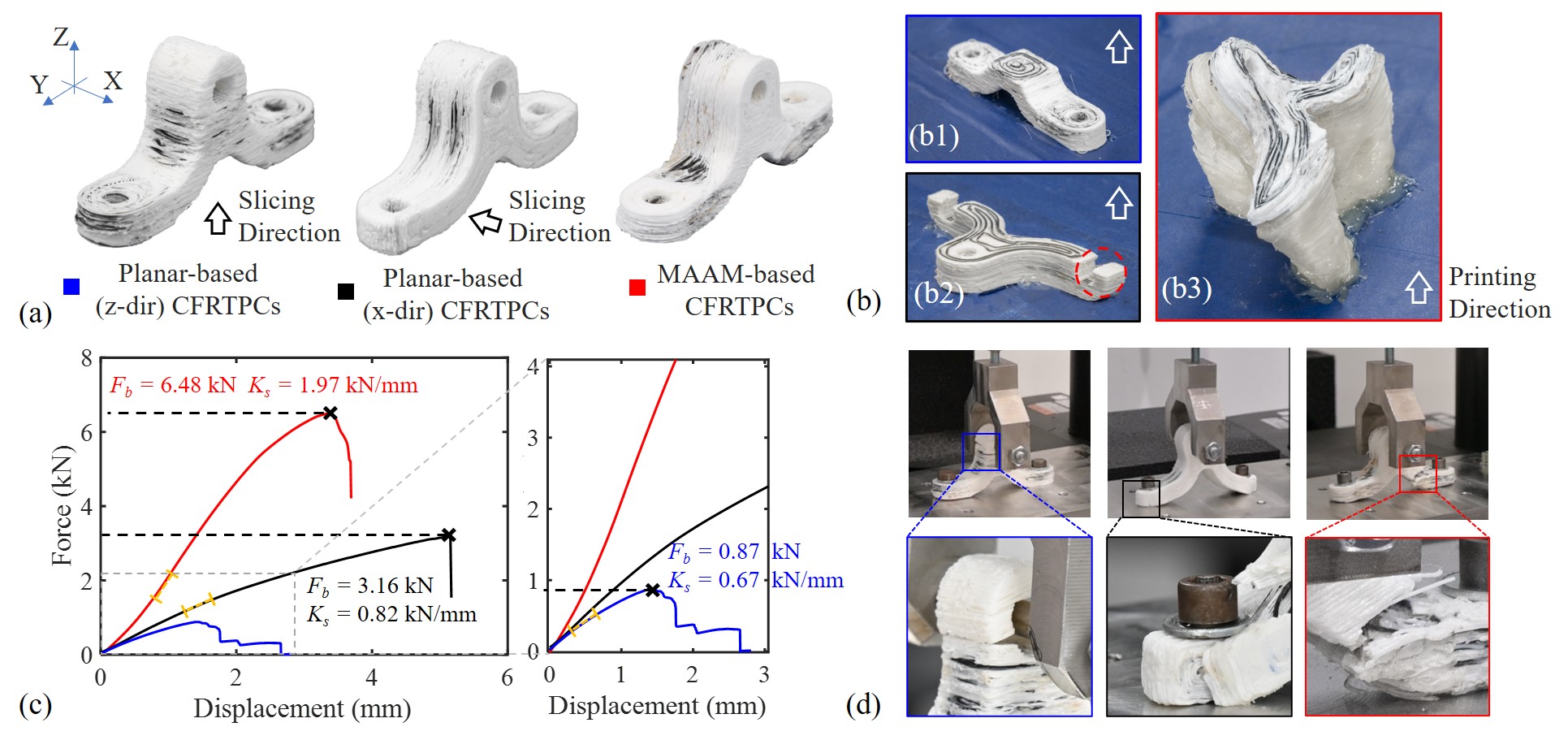}
\caption{Physical experimental results for the T-Bracket model. (a) Fabrication results by using both planar-layer based toolpaths (along two different slicing directions) and our spatial toolpaths for MAAM. (b1-b3) Printing processes when employing different strategies. (c) Tensile test results -- the model printed using our spatial toolpaths exhibits significant enhancement in both the tensile strength and the \rev{breaking force}{failure load}, proving exceptional reinforcement performance. (d) Models fabricated by different printing strategies also result in different modes of breakages on the specimens in tensile tests.}
\label{fig:TBracketResult}
\end{figure*}

\subsection{Mechanical Performance}
To verify the mechanical performance of the printed CFRTPCs, tensile tests were conducted by using the Instron 5969 Dual Column Testing System which has a working range of $30~kN$. \rev{}{For all experiments, the crosshead speed is set at $2 \mathrm{mm}/\mathrm{min}$.} We tested and compared the performance of models fabricated by both the planar-based strategy and our spatial toolpaths with MAAM. Here, we primarily focused on two measurements: the \rev{breaking force}{failure load} ($F_b$) and the model stiffness (\rev{}{$K_s$} -- calculated as the ratio between force and displacement). The results of the tensile tests for all models are summarized in Table~\ref{tab:comparisonangleave}.

First of all, the substantial enhancement and the exceptional mechanical performance of CFRTPCs by using our method were demonstrated on the GE-Bracket model (Fig.~\ref{fig:teaser}) and the Top-Opt model (Fig.~\ref{fig:topoptResult}). The CCF reinforcement compared to the planar-based print using pure thermoplastic material is noteworthy -- $1587\%$ and $849\%$ enhancements in \rev{breaking force}{failure load} can be achieved on these two models respectively. It is interesting to find that the performance of CCF reinforced model using planar layers is even worse than that purely using thermoplastic material but optimized spatial toolpaths. \rev{}{When the PLA matrix material is aligned with the stress distribution through a spatial toolpath, it demonstrates enhanced resistance to layer delamination, leading to a higher failure load, as presented in our previous work~\cite{fang2020reinforced}. It is important to note that, unlike continuous fiber, the anisotropy in PLA is not inherent to the material itself but results from the bonding between filaments.} The limited effectiveness of the conventional two-and-a-half dimensional fiber placement strategy was well demonstrated by this model 
with 3D stress distribution. Compared to the model fabricated with pure matrix material by MAAM, adding continuous carbon fibers only increases the weight of the model by $17\%$ yet obtains an enhancement of $336\%$ and $54.5\%$ in the \rev{breaking force}{failure load} and the model Stiffness respectively. In short, high mechanical performance can be realized on models reinforced with continuous fibers but only when they are effectively placed to follow the optimized toolpaths.

The fabrication results of the Bike Fork model are given in Fig.~\ref{fig:forkFabrication} together with the tensile test results. It can be observed that the model reinforced with spatially aligned continuous fibers \rev{achieved a breaking force of}{breaks at} 9.27 kN, which is $168\%$ increased compared to the model reinforced by using fibers on planar layers. This significant difference in \rev{breaking force}{failure load} is mainly driven by the fact that fibers fail to continuously connect critical regions when using planar-based layers (i.e., fibers placed at the bolt joint region and the two legs are disconnected). As depicted in Fig.~\ref{fig:forkResult}(c) and (d), material failure exhibits different patterns.  Specifically, the model printed using planar layers broke due to layer delamination. The interface between the matrix material and the fibers was stretched by out-of-plane tensile forces. In contrast, as shown in Fig.~\ref{fig:forkResult}(d), the spatial toolpaths for MAAM can control the breaking region exhibited on the region with the highest stress (i.e., the leg region, as disciplined by FEA result in Fig.~\ref{fig:forkResult}(a)), which leads to the final breakage of fiber fracture. The advantage of spatial alignment of fibers is also evident in the enhancement of model stiffness. In this case, \rev{}{the model stiffness $K_s$} has been enhanced by $73.2\%$.

From the tensile test results of the T-Bracket model given in Fig.\ref{fig:TBracketResult}, it is found that the printing direction can remarkably influence the model's strength even when utilizing the planar-based printing strategy. When the model is printed along the x-axis using 2.5D toolpaths, it demonstrates a \rev{breaking force}{failure load} (3.16 kN) over three times the model printed along the z-axis direction (0.87 kN). Although this model presents relatively planar structures, the CFRPTC fabricated using MAAM with spatial toolpaths still showcases the highest mechanical performance. This is because the fibers can align successfully across load-bearing regions that are perpendicular to each other. By perfectly aligning with the 3D stress distribution, the MAAM-based reinforcement strategy achieves a \rev{breaking force}{failure load} level of $F_b = 6.48 kN$ (as depicted in Fig.\ref{fig:framework}(b)). This represents an enhancement of $105.1\%$ and $653.4\%$ over the planar-based models sliced along the z-direction and x-direction respectively. Furthermore, the breaking regions under the same loading vary according to different fiber placement strategies as illustrated in Fig.\ref{fig:TBracketResult}(d). In the scenario of z-direction planar-based printing, fractures occur in the head region where the fiber provides less assistance in preventing layer delamination. For x-direction planar-based printing, small patches on the bottom corner (highlighted in a red circle in Fig.\ref{fig:TBracketResult}(b2)) prevent the placement of continuous fibers. Only with MAAM can the fibers be arranged effectively across the entire design domain, facilitating the most robust method of reinforcement and resulting in direct fiber breakage at the most critical region (i.e., the leg region). Aligning continuous fibers long the spatial toolpaths also guarantees the achievement of the highest model Stiffness, showcasing a $140.2\%$ and $194.1\%$ enhancement compared to the other two planar-based results.

\subsection{Discussion}\label{subsec:discussion}
In this work, while exceptional mechanical performance has been achieved on models with 3D stress distribution through effective fiber reinforcement via spatial toolpaths printed by the MAAM process, there remains \rev{significant}{certain} potential for further improvement. \rev{In the following discussion}{Specifically}, we \rev{will explore}{discuss} the \rev{}{application scope of spatial toolpaths, examine the influence of mesh density, and discuss the limitations of our computational pipeline in detail below.}

\vspace{5px} 

\rev{}{(1) \textbf{Advantages and application scope of spatial fiber toolpath:} It is important to highlight that the benefits of implementing spatial toolpaths with MAAM in CFRPTCs fabrication are not only for the models with complex geometry. Even for models with relatively simple geometry, such as the T-bracket and Twist Bar models featured in this work, they can exhibit 3D stress distribution under certain loading conditions such as shown in Fig.~\ref{fig:framework}(b) and Fig.~\ref{fig:barModelResult}(b). In these scenarios, traditional planar-based slicing and toolpaths may not provide adequate reinforcement. On the other hand, our computational pipeline as a general framework can also generate effective planar toolpaths for models with predominantly 2D stress distribution. 
The most significant contribution of our approach is evident in models with 3D stress distributions. This advantage can be quantitatively assessed by analyzing the angle distribution between the stress direction and a fixed printing orientation. When there is a substantial misalignment between these two directions as can be commonly observed in planar-based models (demonstrated in Fig.~\ref{fig:PSAlignAngle}), our spatial toolpath solution is particularly effective for enhancing the mechanical strength.}

\vspace{5px}

\begin{figure}[t]
 \centering 
\includegraphics[width=\linewidth]{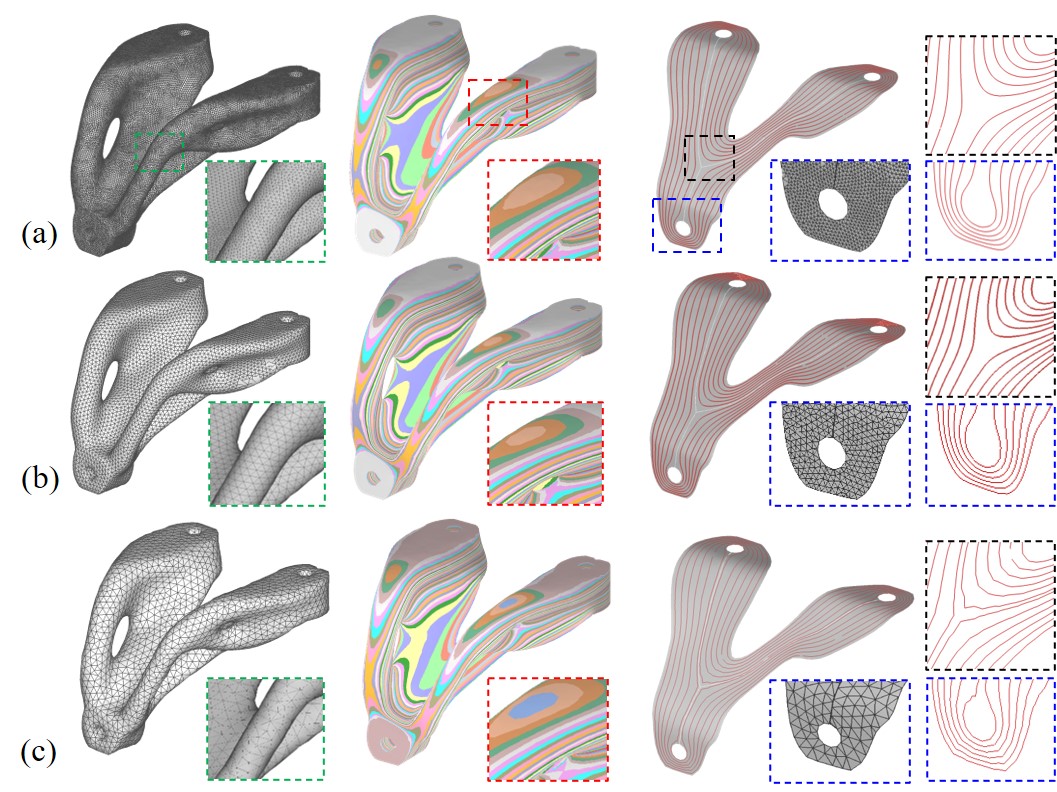}
\caption{\rev{}{Comparison of the computational result based on different mesh densities. (a) Dense mesh with an average edge length of $l_{e} = 1.21~\mathrm{mm}$. (b) The mesh density used in our implementation, featuring an average edge length of $l_e = 2.52~\mathrm{mm}$. (c) Results computed on a sparse mesh with an average edge length of $l_e = 5.28~\mathrm{mm}$.}
}\label{fig:meshDensity}
\end{figure}

\rev{}{(2) \textbf{Influence by mesh density.} As outlined in Sec.\ref{subsec:compResult}, the selection of mesh density is based on the model's size and isotropic tetrahedral mesh is used in this work. Here we discuss the impact of mesh density on both curved layer slicing and toolpath generation. The second column of Fig.\ref{fig:meshDensity} demonstrates that our approach can yield similar results in curved layers across different resolutions -- as long as the mesh is sufficiently fine to capture the geometric details of an input model. However, a sparse mesh can result in a non-smooth toolpath (as shown on the right of Fig.~\ref{fig:meshDensity}(c)). This is because a sparse mesh provides limited computational space for generating a smooth guidance field, leading to fiber toolpaths that do not effectively satisfy the objectives. However, the mesh refinement process can always be invited to resolve this issue. Additionally, for large models with complex geometries such as those generated through topology optimization, advanced mesh generation methods like adaptive variable density mesh~\cite{ni2018field} could be employed to improve the efficiency of computation.}

\vspace{5px} 

(3) \textbf{Drawback of low fiber-to-volume ratio.} This is a key direction for future development and can further enhance the mechanical strength of printed CFRPTCs with spatial toolpaths. Based on the SEM image \rev{}{of fiber-dense regions}, as shown in Fig.~\ref{fig:GEBFabResult}(e), fibers cover approximately $26\%$ of the area. Considering the $34.5\%$ dry fiber ratio of the pre-impregnated CF-FR-50 material (documented in the Markforged datasheet~\cite{Eiger}) and the incomplete fiber filling in curved layers, the final fiber volume fraction in printed CFRPTCs is less than $5.0\%$. \rev{}{Additionally, considering that CF-FR-50 is a material embedding continuous fiber within a Polyamide (PA) matrix, the weak bonding between the PLA matrix and PA could also contribute to potential weaknesses in material performance.} 
Despite using spatial fiber toolpath leading to an average improvement of over $300\%$ in strength compared to models made solely of matrix material, there's large room for future enhancements. This could be achieved by either generating denser fiber toolpaths on curved layers using different patterns (e.g., stripe pattern reported in~\cite{knoppel2015stripe}) or adopting an in-nozzle impregnation method~\cite{Matsuzaki2016_SciReport, Li2020_CompositeB} that yields a higher fiber-to-volume ratio.

\vspace{5px} 

\rev{}{(4) \textbf{Limitation of our computational algorithm.}}
\rev{}{The computational pipeline developed in this work for spatial toolpath generation has certain limitations in a few steps. For instance, our stress analysis and continuity protection techniques have been primarily demonstrated on models featuring bolt joints. Other geometry features and loading cases, such as compression, have not been explored. This can be a crucial area for future exploration. Additionally, the toolpath computing algorithm relies on topology analysis and segmentation for curved working surfaces. While this process is generally effective for curved surfaces with complex topology (e.g., layer 22 of the GE-bracket model illustrated in Fig.\ref{fig:teaser}(d)), it encounters challenges in cases where multiple holes experience loading conditions that are not perpendicular to their boundaries. In extreme scenarios, different objectives such as $\mathcal{O}_{sf}$ and $\mathcal{O}_{cp}$ can become contradictory with each other, thus leading to toolpaths that deviate significantly from the distribution of principal stresses. More sophisticated methods need to be developed to enhance the capability of our topology analysis step to handle more complicated loading conditions.}

\section{Conclusion \rev{}{and Future Work}}~\label{Sec:Conclusion}
We demonstrated the functionality of multi-axis additive manufacturing for fabricating CFRTPCs with spatially placed fibers, which can realize exceptional mechanical performance. An effective computational pipeline has been developed to generate optimized spatial fiber toolpaths on given models with loading conditions, satisfying two critical requirements: following the maximal stress directions in critical regions and connecting multiple load-bearing regions by continuous fibers. The computed toolpaths have been successfully employed to fabricate CFRTPCs reinforced by spatially aligned fibers by using a system with dual robotic arms. The performance of reinforcement has been verified and compared with the planar-layer based method via tensile tests. Across all examples presented in this paper, we observe an enhancement varying from $105.1\%-544.0\%$ and $59.5\%-140.2\%$ in the \rev{breaking force}{failure load} and the model Stiffness respectively. This research marks the first attempt at 3D volumetric decomposition and multi-axis additive manufacturing for continuous fiber-reinforced composites, paving the way for maximizing the effectiveness of fiber reinforcement.

\rev{}{In this paper, we mainly focus on the computation within a fixed design space, and a promising }strategy to further enhance the mechanical strength of CFRTPCs fabricated by MAAM is to take the co-optimization principles (i.e., design for additive manufacturing~\cite{gao2015status}). Despite advancements achieved through curved layer slicing and optimization based pipelines, avoiding small patches in models like the Top-Opt remains a challenging task. It is necessary to integrate the fabrication requirements into the structure design optimization. \rev{}{Moreover, incorporating fiber anisotropy within the loop of optimization stands as a critical task in our future research,} where we could study existing works in the field of topology optimization~\cite{Li2020_CompositeB,jantos2020topology}. \rev{}{On the other hand, the spatial printing is empirically set at a relatively low speed in physical experiments (i.e., robot arm task space speed at 8~mm/min). This results in a longer fabrication time compared with the conventional planar-based CFRPTCs AM process. To enhance the efficiency of MAAM with CFRPTCs, we plan to explore the development of advanced motion planning techniques and toolpath generation algorithms that consider dynamics in the future.}



\section*{Acknowledgments}
The project is partially supported by the chair professorship fund at the University of Manchester and UK Engineering and Physical Sciences Research Council (EPSRC) Fellowship Grant (Ref. \#: EP/X032213/1).

\section*{References}

\bibliography{ref}

\end{document}